\documentclass{aastex}
\usepackage{emulateapj5}
\usepackage{apjfonts}
\usepackage{onecolfloat}
\usepackage{graphicx} 

\newcommand{\etal}{{et~al.}}
\newcommand{\teff}{T$_{\rm eff}$}
\newcommand{\mfeh}{\rm{[m/H]}}
\newcommand{\mteff}{\rm T_{eff}}
\newcommand{\logg}{log~$g$}
\newcommand{\mlogg}{\rm{log}g}
\newcommand{\moo}{O$_2$}
\newcommand{\opof}{[O$_{\rm p}$/O$_{\rm f}$]}
\newcommand{\dT}{$\Delta$T$_{\rm eff}$}
\newcommand{\mV}{M$_{\rm V}$}
\newcommand{\msun}{M$_{\odot}$}
\shorttitle{Oxygen Abundances in Metal-Poor Stars}
\shortauthors{Fulbright \& Johnson}
\begin{document}
\title{Oxygen Abundances in Metal-Poor Stars}

\author{Jon. P. Fulbright\altaffilmark{1,2,3}}

\affil{Dominion Astrophysical Observatory, Herzberg Institute of Astrophysics, National Research Council of Canada, 5071 W. Saanich Road, Victoria, BC V9E-2E7}
\email{jfulb@ociw.edu}

\and

\author{Jennifer A. Johnson\altaffilmark{4}}

\affil{Observatories of the Carnegie Institution of Washington, 813 Santa Barbara St., Pasadena, CA 91101}
\email{jennifer@ociw.edu}

\altaffiltext{1}{Visiting Astronomer, Canada-France-Hawaii Telescope, which is operated by the National Research Council of Canada, the Centre National de la Recherche Scientifique, and the University of Hawaii.}
\altaffiltext{2}{Visiting Astronomer, Kitt Peak National Observatory, operated by the Association of Universities for Research in Astronomy, Inc., under contract with the National Science Foundation.}
\altaffiltext{3}{Present address:  Observatories of the Carnegie Institution of Washington, 813 Santa Barbara St., Pasadena, CA 91101}
\altaffiltext{4}{Present address:  Dominion Astrophysical Observatory, Herzberg Institute of Astrophysics, National Research Council of Canada, 5071 W. Saanich Road, Victoria, BC V9E-2E7}

\begin{abstract}
We present oxygen abundances derived from both the permitted and forbidden
oxygen lines for 55 subgiants and giants with ${\rm [Fe/H]}$ values between $-2.7$ and solar with the goal of
understanding the discrepancy in the derived abundances.  A first
attempt, using \teff{} values from photometric calibrations and surface
gravities from luminosities, obtained agreement between the indicators
for turn-off stars, but the disagreement was large for evolved stars.  We
find that the difference in the oxygen abundances derived from the permitted
and forbidden lines is most strongly affected by \teff{}, and we derive
a new \teff{} scale based on forcing the two sets of
lines to give the same oxygen abundances.  These new parameters, however,
do not agree with other observables, such as theoretical isochrones or
Balmer-line profile based \teff{} determinations.  Our
analysis finds that one-dimensional, LTE analyses (with published 
NLTE corrections
for the permitted lines) cannot fully resolve the disagreement in the two
indicators without adopting a temperature scale incompatible with other
temperature indicators.  We also find no evidence of circumstellar 
emission in the forbidden lines, removing such emission as a possible
cause for the discrepancy.
\end{abstract}


\section{Introduction}
Oxygen is the third most common element in the Universe. It is copiously
produced when massive stars explode as Type~II supernova. This
distinguishes it from Fe, which is also made in Type~Ia SN,
the accretion-induced explosions of white dwarfs. The [O/Fe] ratio 
therefore reflects the mix of stars that have contributed to 
the enrichment of a system. It has been used to diagnose 
the source of metals in 
X-ray gas in galaxies \citep{g97,x02} 
and in damped Ly$\alpha$ systems \citep{p02}.
Because Type~II SN begin to explode more quickly than Type~Ia SN
after stars are formed, the
O/Fe ratio after star formation begins is large at first, 
then declines as Fe, but little O,
is contributed by the Type~Ia SNe \citep{t79}. This fact has been
exploited to argue that Bulge formation lasted $<$ 1 Gyr 
\citep{mr99} and star formation for dwarf galaxies happened in bursts 
\citep{gw91,s02}.  The fact that the oldest stars in our Galaxy 
have supersolar [O/Fe] ratios must be considered when measuring the ages
of globular clusters \citep{v85}.

In particular,
the [O/Fe] ratios in metal-poor stars in the Milky Way are important
because they provide a look at the chemical evolution
of the early Galaxy. We can use the O and Fe abundances to derive
yields from Type~II SNe, to adopt the correct isochrones for globular
clusters, and to calculate the
timescale for the formation of the halo. The [O/Fe] ratios in 
old Milky Way stars also provide a starting point for interpreting
the abundances seen in high-redshift systems. 

Unfortunately, the lines available in late-type stars are not
ideal abundance indicators. The strength of the forbidden lines at 6300 \AA{} 
and 6363 \AA{} are gravity-dependent and are very weak in dwarfs and subgiants.
The triplet of permitted lines at 7771-7774 \AA{} have excitation
potentials of 9.14 eV and therefore are weak in cool giants.  For some
evolutionary stages the
permitted lines are also affected by NLTE effects \citep{k91,g99,m00,nlte}.
The OH lines in the ultraviolet and infrared regions of the spectrum are
 measurable in dwarfs and subgiants.  However,
OH is a trace species in these stars, and is particularly sensitive to
inhomogeneities in temperature \citep{ag01}. 

Many studies using these abundance indicators show
disagreement in the [O/Fe] vs. [Fe/H] relationship for 
stars with [Fe/H] $< -1.0$ (see Figure 1 for an incomplete, but
demonstrative, summary). Because [O~I] lines are stronger in giants and
O~I lines in dwarfs, studies using different indicators also 
use data from different
types of stars. In general, the studies using permitted O~I 
lines \citep{ar89,t92,cv97} and the UV OH lines 
\citep{i98,i01} in dwarfs and subgiants 
find a steep linear increase in [O/Fe] with 
decreasing [Fe/H].  \citet{b99} combined O~I and UV OH measurements 
and found a slope of $-0.35$. In contrast, the [O~I] lines in giants
and subgiants give [O/Fe] values 
that plateau at $+0.35$ for [Fe/H]$<-1.7$ \citep{g86,b88}. 
More recent analyses \citep{k00,s01} show instead a slight slope, but a 
difference of $\sim 0.5$ dex between the indicators at 
[Fe/H] $= -3.0$ remains.  The O abundances measured from the infrared OH lines 
in dwarfs, subgiants, and giants 
produce similar values to the [O~I] lines \citep{b01,m00}.  

It is possible that the differences cited above are the result of intrinsic
variations in the oxygen abundance between giants and dwarfs. However, 
studies of small samples of dwarfs with $-2.0 < $[Fe/H] $<-0.5$ 
(\citealt{ss91}, 7 stars; \citealt{sw91}, 2 stars) showed 
that the [O~I] line in these stars gave an oxygen abundance 0.4-0.7 
dex lower
than that derived from the permitted lines in the same stellar spectra.
Thus the discrepancy between forbidden and permitted lines cannot be
ascribed alone to different intrinsic oxygen abundances in giants and dwarfs.

There have been many attempts to find another solution and to 
reconcile the results produced by the
different sets of lines, 
either through finding the same slope and intercept in the
[O/Fe] vs. [Fe/H] relation for different samples of stars or through 
finding the same O abundance using different lines in the same star. 
Oxygen abundances 
are sensitive to the adopted stellar parameters, so several studies have
argued for improved methods for finding the parameters.
\citet{k93} constructed new color-\teff{} scales
that produced 
effective temperatures that were 150--200 K hotter than those used
by other investigators.  These higher temperatures decreased the
derived O abundance from the permitted lines so that they gave
the same [O/Fe] ($\sim 0.5$ dex) at low metallicities seen in giants.
\citet{cv97} also found that temperatures that were hotter by 150 K than
their original temperature scale would erase the discrepancy in five turnoff
dwarfs and subgiants with [Fe/H] $< -1.0$.

Recently, the gravities, rather than the temperatures, have come
under scrutiny.  \citet{k00} re-evaluated the [O/Fe] values for metal-poor 
dwarfs from \citet{b99}  and \citet{t92},
in light of NLTE effects on Fe~I \citep{ti99}. King adopted
gravities from \citet{ti99} and \citet{ax95} which were based on Fe~I/Fe~II
ionization balance, but with NLTE corrections included for Fe~I,
and based the [Fe/H] scale on Fe~II instead of Fe~I.  When this is done,
the O~I abundances show the same slight slope as the [O~I] abundances, though
they were still higher. For five unevolved stars with both [O~I] and O~I
measurements, the O~I-based abundances exceeded the [O~I] by 
$+0.24\pm 0.05$ dex. 
\citet{cg00} analyzed 40 stars (7 with [Fe/H] $< -1$) with measured
O~I and [O~I] lines, ranging from dwarfs to giants.  The O~I
abundances were corrected for NLTE effects using the results of
\citet{g99}, and they observed no difference between the two
indicators on average, with the exception of the cool giants. The 
tendency of the 
permitted lines of giants to give higher abundances than the forbidden
was attributed to deficiencies in the \citet{kz92} models that were
used in the analysis.

\citet{n02} obtained high-resolution, very high S/N ($>$ 400)
data for 18 dwarfs and subgiants with $ -2.7 < $[Fe/H] $< -0.5$.  Their
equivalent width measurements have errors of $< 0.3$ m\AA{} for the forbidden
and $< 1$ m\AA{} for the permitted lines. The quality of their data allowed
the forbidden lines to be measured in higher-gravity metal-poor stars than
before. When they used 1-D model atmospheres and NLTE corrections, 
the [O~I], O~I triplet and
UV OH lines gave the same [O/Fe] vs. [Fe/H] relation. 
However, consideration of 3-D effects, in
particular granulation, only reduced the oxygen abundance derived from the
[O~I] lines, and a disagreement remained at the level of 0.3 dex.
\citet{n02} compared their [O/Fe] values in dwarfs with
those in giants of the same metallicity. While the [O~I] lines gave
satisfactory agreement, the O~I triplet lines in giants gave higher abundances
than those seen in dwarfs and subgiants.

One metal-poor subgiant, BD~+23~3130, has been subjected to intense 
scrutiny by several authors. 
\citet{i98} found 
[O/Fe] $= +1.17 \pm 0.40$ for this star using the UV OH lines and O~I triplet. 
\citet{fk99} argued that this was incompatible with the weakness of 
the [O~I] line at 6300 \AA{}, which yielded [O/Fe] $= +0.35 \pm 0.20$.
\citet{c01} observed the 6300 \AA{} line of this star at S/N $\sim$ 900 and 
measured an equivalent width of 1.5$\pm$0.5m\AA{} and found  
[O/Fe]$=0.71 \pm 0.25$, halfway between the \citet{i98}
and \citet{fk99} values.  \citet{i01} revised the 
analysis using a \logg{} value 0.4 dex higher than their previous study.  
With this
analysis, they found agreement among the UV lines, 
the [O~I] line and the O~I triplet. 
\citet{n02} used similar atmospheric parameters, but OSMARCS
models also achieved agreement between the [O~I] and UV OH lines with 1-D
atmospheres, but not 3-D atmospheres.  The studies of \citet{n02} and 
\citet{i01} suggest that a solution may be found in the application of correct 
stellar parameters and a consistent analysis of Fe and O using those parameters.

While using different indicators for different samples of stars increases
the number of possible targets, especially at low metallicity, we will be
looking instead at a sample that have both [O~I] and O~I lines.
Using both sets of lines in the same star is important because  
star-to-star variations exist for oxygen and other element-to-iron ratios in 
metal-poor stars \citep{c97,k97,h98,f02}.  The most glaring example of this is
the subgiant BD +80 245, whose permitted O~I lines give a sub-solar [O/Fe]
ratio at [Fe/H] $\approx -2.0$ \citep{c97}. Also, such a tactic 
avoids the question of whether oxygen has been depleted by internal mixing
in giants, which means that they can be included in the sample.
Therefore, a more rigorous way to insure both the permitted and
forbidden oxygen lines truly give the same results is to use both lines
in the same stars.  

The recent studies of \citet{i01} and \citet{n02} showed that agreement
between the oxygen abundance given by O~I and [O~I] could be reached, at
least for turnoff dwarfs and subgiants, for their particular choices
of 1-D atmospheres. However, because of the weakness of the [O~I] line
in dwarfs and subgiants, there are only six stars in these two papers that
have both [O~I] and O~I measurements. We chose to focus on subgiants and
giants to obtain a large, homogeneous sample of stars with both sets of
lines measured, including a number with [Fe/H] $< -1.5$. This will also test
whether the successes with the dwarfs and subgiants can be replicated, or
whether, like \citet{cg00}, the analysis of cool giants will produce 
different oxygen abundances.
We have taken advantage of the very high resolution (R $\sim 130000$) 
Gecko spectrograph on CFHT to obtain equivalent widths for the [O~I] lines 
for a sample of 55 stars, mostly subgiants and giants,
 with [Fe/H] between $-$2.7 and solar.   Additional 
spectra and literature sources have been included so that all 55 stars have 
measurements of both the permitted and forbidden oxygen lines.

The data set presented here provides a strong test of any attempts
to reconcile the indicators. 
We will begin our analysis by measuring the magnitude of the
difference in these two oxygen abundance indicators when we adopt atmosphere
parameters with \teff{} from colors and \logg{} from isochrones. We find that the
familiar pattern of O~I lines giving higher O abundances than the [O~I] lines
reasserts itself. Next, we
examine whether
changing the assumptions of the analysis, in particular the temperature scale, 
eliminates the measured difference.  We then use the knowledge of the 
behavior of the lines to create an ad hoc parameter set that, within
the assumptions of the analysis, results in agreement between the
indicators. Finally, we discuss whether this ad hoc parameter
scale is realistic when compared to other observables for the target stars.

\section{Methodology}
 
An exhaustive study of all the possible solutions to the oxygen abundance
problem is beyond the scope of a single paper. We therefore
concentrate on following up the apparent successes of parameter-based
solutions in the recent works mentioned in the Introduction.  
 
Our analysis follows the following assumptions:
 
1) The atmospheres of stars can be described by one-dimensional,
plane-parallel models in LTE. For most of this work, we will use
Kurucz (1995) models\footnote{Available from http://cfaku5.harvard.edu/}. 
We use the MOOG stellar abundance package \citep{s73} for the analysis.
We adopt $\log{\rm n(Fe)_\odot} = 7.52$ and $\log{\rm n(O)_\odot} = 8.69$. 
The later value is based on the reanalysis of the solar [O~I] by 
\citet{ap01}, which
takes in account the contamination of the 6300 \AA{} [O~I] line by 
a Ni I weak line.

2) Non-LTE effects limit the usefulness of Fe~I lines \citep{ti99}
in metal-poor stars.  Non-LTE conditions also affect the permitted O~I lines, 
but for the purposes of this experiment we will assume the abundance 
corrections of \citet{nlte} adequately compensate
for the departures from LTE.  We also assume the lines of Fe~II are free of 
non-LTE effects and will be used as the primary Fe abundance indicator.

3) Within the assumptions above, we will assume the solution to the problem 
can be found
by the application of the correct atmospheric parameters for the stars.
This assumption is similar to the solution put forth by \citet{k00}.

The methods employed here are similar to those taken by 
\citet{n02} and \citet{k00}, but their samples contain
 only warm (\teff $> 5600$ K) 
turn-off and subgiant stars, and only have 11 stars between them
with both forbidden
and permitted lines.

\section{Target Star Selection and Observations}

The Gecko echelle spectrograph on the Canada-France-Hawaii Telescope (CFHT) 
delivers spectra with resolution of $\sim 130000$, with a dispersion of
$0.018$ \AA{} per 13.5 $\mu$m pixel. Only one spectral
order is observable at a time (selected by a narrow-band filter), meaning 
only a $\sim 75$ \AA{} region is observed per exposure with the 
thinned 2048x4096 pixel detector. 
The Gecko data were obtained over 6 nights in April and September
2001. In both runs, the spectra covered the wavelength range from
6290--6370 \AA, covering both the 6300 \AA{} and 6363 \AA{} lines. 

The primary candidate list was created in a similar manner as the target
list of \citet{f00}, using literature lists of known metal-poor
stars, such as \citet{bond}, \citet{c94}, etc.  The list was then
culled of stars where we estimated that one of the two sets of lines would 
be undetectable.

The spectra were reduced using normal IRAF\footnote{IRAF is distributed by 
the National Optical Astronomy Observatories, which are operated by the 
Association of Universities for Research in Astronomy, Inc., under cooperative 
agreement with the National Science Foundation.} 
routines.  Although previous work has shown that the scattered light 
effect in Gecko is less than 1\%, special care was taken in its removal.
A wavelength solution was applied using ThAr lamps taken at the beginning
and end of the night. The variation in the wavelength of the telluric \moo{} 
features between spectra is less than 100 m s$^{-1}$. Details of the
individual observations are given in Table 1.

Because the 6300 \AA{} feature lies within a band of telluric \moo{} lines, 
it was sometimes necessary to remove these telluric lines from the spectrum. 
During the observation runs, spectra of bright, rapidly-rotating 
(vsini $> 250$ km s$^{-1}$), 
spectral type B or A stars were observed. These spectra were used to
divide out the telluric \moo{} features (some sample spectra are shown in
the Appendix). For most stars the division was cosmetic, 
because the [O~I] line was not contaminated and the very high resolution
and dispersion of the Gecko spectrograph lessens the probability 
of contamination.   

\citet{l91} suggested circumstellar [O~I] emission may play a 
role in the discrepancy. We did not observe any sign of stellar [O~I]
emission in our spectra, a point which we examine this point further 
in the Appendix.

Additional data to measure the Fe and O~I lines
 were obtained with the Lick 3-m and KPNO 4-m with their
respective echelle spectrographs.  The new data taken with the Hamilton
spectrograph at Lick were obtained in the same way as the previous data 
(see \citealt{f00} and \citealt{j02} for more details).  The echelle data 
from KPNO were
obtained in January 2002 with a resolution of $\sim 40000$ and covers the 
wavelength range from 4480 to 7850 \AA.  Included in the KPNO data is a
spectrum of the asteroid Vesta, which provides a solar spectrum taken
as if the Sun was a point source.

\section{Line Measurement}

The equivalent widths (EW) values of the 6300 and 6363 \AA{} [O~I] lines were 
measured using both Gaussian fits and  
integrations. The EW of the 6300 \AA{} line for BD~+23~3130 was adopted 
from \citet{c01}. The EW values for the permitted O~I lines were 
measured in the non-Gecko data or taken from literature sources or 
a combination of both. Table 2 gives the oxygen EW values for the stars 
analyzed in this paper.  The EW of the 6300 \AA{} [O~I] line
has been corrected for contamination by the 6300.34 \AA{} Ni I line 
(see Section 6.1).

We adopt the \citet{l78} $gf$-values for the 6300 and 6363 \AA{}  
forbidden lines ($\log{gf} = -9.75$ and $-10.25$, respectively), and
the \citet{b90} $gf$-values for the 7772, 7774, and 7775 \AA{} 
permitted lines ($\log{gf} = +0.36$, $+0.21$, and $-0.01$, respectively).
The Fe line list of \citet{f00} is used here.  
The atomic data for the Fe~I lines are from \citet{obr} and the Oxford
group \citep[and references therein]{oxf}, while the  Fe~II lines have 
data from \citet{blw80} and \citet{moity}.  Slight modifications have
been made to the $gf$-values to improve consistency between sources, 
as described in detail by \citet{f00}.
Many of the target stars have been analyzed by the authors before 
\citep{f00,j02}. We adopt the Fe EW values from those papers. The Fe EW 
values for the previously-unpublished stars are given in Table 3a and 3b 
(available in the electronic version only). 

The oxygen abundances found for the solar analysis are larger than the adopted
solar oxygen abundance of \citet{ap01} by 0.14 (forbidden) and
0.10 (permitted) dex.  We could change the $gf$-values of the
lines to reflect these differences, but the value of 8.69 comes from a
3-D analysis, which we do not do here.   Allende Prieto et al. report that
using a one-dimensional model would increase the resulting solar oxygen
abundance by 0.08 dex, in reasonable agreement with the solar
abundance derived by the 1-D analysis conducted here.  Therefore, we
choose not to do a differential abundance analysis.

For this paper, the ratio of the abundances given by the two oxygen indicators
is more important than the absolute abundance.  Using the present $gf$-values,
the solar analysis yields a forbidden line oxygen abundance 0.04 dex larger 
than that obtained permitted lines.  However, if
we assume that all of the uncertainty is from line measurements error, we get
an uncertainty for the ratio of 0.06 dex (dominated by the uncertainty
in the EW of the 6300 \AA{} line).  Therefore, we believe that a change in
the $gf$-values is not warranted by the analysis.  If the $gf$-values were 
changed to force agreement, the needed \dT{} vaules in Section 7 would be 
increased by about 35 K.

\section{Stellar Parameters}

\subsection{Effective Temperatures}

Initially, we will analyze the oxygen and iron
abundances using stellar parameters derived from two photometric temperature 
scales:  Alonso et al. scale (1996 for dwarfs and 1999 for giants) 
and \citet{h00}. The Alonso scales are based on the Infrared Flux 
Method (IRFM) of \citet{b79}, while the Houdashelt scale is based on 
synthetic spectra with zero points based on observations. 

The input photometry for
these relationships came from a variety of literature sources. The
B$-$V and V$-$I data are from the Hipparcos/Tycho catalog, and
the ubyv$\beta$ photometry is from \citet{hauck}. 
K colors were taken from the papers of Alonso et al. (1994 and 1999), 
\citet{c83}, \citet{l88}, and the 2MASS Point Source 
Catalog. Many V-R colors were taken from \citet{rem}, while
others come from \citet{l88} and \citet{c83}.

Measurements of the reddening were taken from literature sources such 
as \citet{att} and \citet{c94}.  Other reddenings were derived using 
ubvy$\beta$ photometry and the calibration of \citet{s89}, 
although the limits of that calibration exclude many evolved stars. We 
adopt E(B$-$V) $= 1.37$ E(b$-$y) and the 
transformations of \citet{r85}. For the 8 stars for
which we could not find or derive reddening estimates, we assume zero 
reddening. 

The final dereddened colors are given in Table 4, while the calculated
and adopted \teff{} values are given in Tables 5 and 6. 
In all cases the \teff{} values were only accepted if the star's parameters 
were within the limits of a given color's calibration.
Because both Alonso and Houdashelt give different calibrations for giants and
dwarfs, stars with derived \logg{} $> 3.5$ were considered dwarfs
while the remaining stars were considered giants. While we initially 
intended to adopt the mean \teff{} value for the analysis, the
\teff{} values derived from the (V$-$I) colors were consistently higher.
Additionally, the spread between the results for the different \teff$-$color 
relations for individual stars was sometimes very large, most likely due 
to problems with the photometric data.   
Therefore we ignored most of the results from the (V$-$I) \teff$-$color relation 
and other discrepant points when deciding which \teff{} value to adopt for
each star.  
The final values have been rounded to the nearest 25 K increment.
For convenience, we list the final \logg{}, [m/H], and
$v_t$ values for each star in Tables 5 and 6.  

We have assigned a measure of the uncertainty in \teff{} to each star.
In most cases, the value is the standard deviation of the \teff{} values 
used in the final calculation of the adopted value.  While the agreement
between the individual \teff$-$color relationships for some  stars is
quite good, we believe that the uncertainty in the photometric data and
the calibrations of the \teff$-$color relationships place a lower limit
of 75 K on the \teff{} uncertainty.

\subsection{Surface Gravities}

Many traditional abundance analyses derive surface gravities from forcing
agreement in the abundances derived from the Fe~I and Fe~II. 
\citet[hereafter TI99]{ti99} and \citet{ap99} both 
present evidence that 
the Fe~I lines in very metal-poor stars suffer from non-LTE effects.
Therefore LTE analyses of these lines do not give reliable
abundances. We will derive surface gravities for our
stars from the mass ($M$), absolute V magnitude (M$_{\rm V}$), a bolometric 
correction ($BC$), and effective temperature (\teff):

\begin{eqnarray}
\log{g} = \log{\frac{M}{M_{\odot}}} - 0.4(M^{\odot}_{\rm bol} - M_{\rm V} - BC) + 
\nonumber \\
4\log{\frac{\rm T_{eff}}{\rm T_{eff}^{\odot}}} + \log{g_{\odot}}.
\end{eqnarray}

We adopt ${\rm T_{eff}^{\odot}} = 5770$ K, $\log{g_{\odot}} = 4.44$ and 
$M^{\odot}_{\rm bol} = 4.72$. The adopted stellar masses were based mostly on 
the star's assumed position on the appropriate-metallicity 12 Gyr 
\citet{v00} isochrones (adopting a 10 or 14 Gyr isochrone results
in negligible differences).  However, several stars are likely to
have evolved beyond the first-ascent giant branch and probably have
undergone some form of mass loss. For these stars we adopt 
$M = 0.6$ \msun{} (see below). Bolometric corrections were calculated 
from Alonso \etal{} (\citealt{a95} for dwarfs and \citealt{a99} for giants). 

The adopted \mV{} magnitude, especially for giants, can be fairly 
uncertain.  For stars for whose Hipparcos parallax value has 
$\sigma_{\pi}/\pi < 0.25$, we adopt the Hipparcos \mV{} value. Many 
of the remaining stars, especially the giants, have poor Hipparcos 
parallax determinations.  However, \citet{h98} and \citet{att}
derive \mV{} values for many giants and subgiants.  \citet{h98} used
Hipparcos parallax data to improve the \mV{} values derived by \citet{bond},
which themselves were based on fits to globular cluster color-magnitude 
diagrams.  \citet{att} derived distances using Str\"omgren photometry and
\citet{n85} relationships between \mV{}, [Fe/H] and color.

For all the 
non-horizontal branch (HB) stars, we also derived estimates for the \mV{} 
value by using their dereddened colors to place them on the 12-Gyr \citet{v00} 
isochrone appropriate 
for their estimated [Fe/H] value. For stars with estimated [Fe/H] values 
lower than the $-2.31$ limit of the isochrone grid, the [Fe/H] $= -2.31$ 
isochrone was used. 

A number of the target stars are HB or
asymptotic giant branch (AGB) stars, which affects their adopted \mV{}
and mass values.   Following \citet{e97}, we identify 
potential post-RGB candidates in the distance-independent $c_0$ vs. $(b-y)_0$ 
plane (see Figure 2). The locus of subgiant and first-ascent
giants is traced by ubyv isochrones kindly provided by Clem \& Vandenberg
(private communication). 

For the stars that fell into the HB or
AGB regions of the diagram, we adopt a mass of 0.6 \msun{} following
\citet{g98}. The adopted \mV{} value for the assumed HB
stars near the ZAHB locus follows M$_{\rm V}^{\rm HB}=0.19($[Fe/H]$+1.5)+0.61$ 
\citep{g98}.
For the evolved HB and AGB stars, the method of determining the \mV{} 
value was no different than other stars, although a lower limit to the
\mV{} value was placed by the appropiate M$_{\rm V}^{\rm HB}$ value. 

The \mV{} value for stars without high quality Hipparcos
parallaxes or stars not on the HB was based on a 
combination of the \citet{h98}, \citet{att}
and isochrone-derived values. Table 7 lists the \mV{} values from the 
various sources, as well as the final adopted \mV{} and stellar mass values.
The errors in \mV{} were calculated from the Hipparcos parallax or estimated
from the source papers. For the HB stars, an error of 0.2 mag was adopted
to account for uncertainties in the \mV$-$[Fe/H] calibration and any
evolution above the horizontal branch. A color--absolute magnitude
diagram for the final adopted values is shown in Figure 3.  For reference,
a 0.5 mag error in \mV{} contributes a 0.2 dex uncertainty in \logg.

\subsection{Atmospheric [m/H] and $v_t$ Values}

An estimate of the [Fe/H] value for each star was
taken from the literature source of the star. The adopted atmospheric
[m/H] value\footnote{For clarity, we use [m/H] to denote the adopted
abundance scaling for the model atmosphere, while [Fe/H] is the derived Fe
abundance. While the values are usually similar, because we have adopted 
atmospheres with solar abundance ratios, [m/H] is an input value, while [Fe/H] 
is an output value.} was $\sim 0.1$--$0.2$ dex higher because 
most metal-poor stars
have enhancements in the so-called $\alpha$-elements (O, Mg, Si, Ca, etc.),
which provide more free electrons than are accounted for in the solar
ratio models. After the first abundance analysis iteration of
Fe lines, the [m/H] value was based on the [Fe/H] value 
derived from the Fe~II lines. 

We use the $v_t$ value that gave
a flat distribution of derived Fe~I abundances as a function of line
strength. Errors in the adopted $v_t$ value
have negligible effect on the derived oxygen abundances because most of the
oxygen lines are weak. 

\section{Abundance Analysis}

\subsection{NLTE and Ni corrections}

We use the NLTE corrections of \citet{nlte}. The grid of corrections 
only includes stars warmer than 4500 K,
but our sample includes stars several hundred degrees cooler than that
limit. For stars outside the Takeda et al. grid, we have calculated
corrections using an extrapolation from the nearest grid points.

\citet{g00} and \citet{m00} also derived NLTE
corrections for oxygen lines. A comparison of the Gratton et al. and
Takeda et al. corrections for the 7772 \AA{} O~I line is shown in Figure 4. 
We calculated the comparisions by assuming the same 
O~I line strength using the listed [O/Fe]$_{\rm LTE}$ value for each
metallicity.  The only large difference between the two calculations is for
the hot, low-surface gravity stars. Adopting the Gratton \etal{} correction
for the HB stars in our sample would result in a reduction of the NLTE 
correction by less than 0.15 dex. 

The 6300.31 \AA{} [O~I] line is blended with a Ni I line at 6300.34 \AA{}.
While the Ni line is fairly weak, it does affect the derived oxygen
abundance in the Sun \citep{ap01}. As a correction, we have
subtracted the estimated strength of the Ni I line (calculated using
the adopted Alonso \teff{} scale models and assuming [Ni/Fe] = 0) from 
the measured 6300 \AA{} EW. In general the correction was only a small
fraction ($< 10\%$) of the adopted EW value, only being significant in 
metal-rich stars like the Sun. The EW values for the 6300 \AA{} line given in 
Table 2 reflect the corrected values.

\subsection {Error Analysis}

\subsubsection{Equivalent Width Errors}

The weakness of [O~I] and O~I lines in metal-poor stars
means that EW errors can dominate the error budget and need to be
considered carefully. \citet{c88} gives a useful derivation of the error
in EW. We write the EW as 

\begin{equation}
EW=\delta x \Sigma{} (C_i - r_i),
\end{equation}
where $\delta x$ is the dispersion in \AA/pix, $C_i$ is the value of the 
continuum, and $r_i$ is the intensity at pixel $i$. This is 
summed over the $n$ pixels which contain absorption in the line. In practice, 
we summed over 15 pixels in the Gecko case and 7 or 8 pixels for the KPNO or 
Lick data, respectively.  The error in the EW, taking into account that the
errors in $C_i$ are completely correlated is

\begin{equation}
\delta EW ^2 = \delta x ^2 \left (\Sigma \delta r_i ^2 + n^2 \Sigma \delta C_i ^2 \right ).
\end{equation}

We used the S/N of the 6300 \AA{} region as the measure of $\delta r_i$.
$\delta C_i$ is also based on the S/N, but because
we averaged $\sim$ 50 pixels around
the oxygen lines to locate the continuum, 
the error in continuum is $\delta r_i/\sqrt{50}$.
We determined the S/N by actually measuring the s.d. in the spectra, rather
than relying on the photon statistics, though in practice they were
the same. Using the S/N 
ignores other sources of error, in particular scattered light,
but as discussed in \S3, the Gecko spectrograph set up minimizes the
impact of scattered light.

We can check our calculations of the EW error in two ways. First, for
38 stars, we measured all three O~I permitted lines and then determined
the expected error in the average abundance both by using the errors 
derived from Equation 3 and by calculating the 
standard deviation in the mean for those three lines. There was encouraging
agreement, usually to within 0.02 to 0.03 dex. 
Second, we compared our EWs to previous 
published measurements (Figure 5).  For the 6300~\AA{} line, we find an
average offset of $-2.4$~m\AA{} with a rms scatter of 3.9~m\AA{}. The average
offset for the 6363~\AA{} is 0.0~m\AA{} with an rms scatter of 2.2~m\AA.
A number of \citet{g00} EW values
are higher than the Gecko observations, while the one discordant 6363~\AA{}
value, from BD~+30~2611 (=HIP~73960), has a 6300~\AA{} EW that agrees with 
the \citet{k92}         
value. The ratio between our EW values for the 6300 and 6363~\AA{} lines 
in this star do 
not follow the expected 3-to-1 ratio (37.7~m\AA{} vs. 17.6~m\AA), but the
Gecko spectrum does not show any indication of the source of the error.
The lines joining the EW values derived for
the same star indicate many cases of large scatter (up to 50\%) among studies 
even when our data are not considered. 

The uncertainty in our EW values from our statistical calculation (generally
on the order of 1 m\AA{} or less) cannot explain the rms deviations
seen in Figure 5. 
The most likely cause for much of the scatter seen is the lower quality of
the previous data. The Gecko data presented here are at higher resolution
and dispersion than the previous data, have negligible scattered light, 
and have minimal \moo{} contamination problems.

\subsubsection{Stellar Parameter-Based Errors}

To measure the effects of systematic parameter errors on the 
oxygen and iron abundances, we ran a series of models for each star: 
one with the \teff{} value raised by 200 K, 
one with the \logg{} value raised by 0.3 dex, 
one with the [m/H] value raised by 0.3 dex, and 
one with the $v_t$ value raised by 0.3 km s$^{-1}$. 
The abundances from each of these 
individual runs were compared against the results of the original run.

The overall effect of the parameter changes on the whole sample
of stars is given in Table 8.  Both the permitted and forbidden lines
are affected by the \teff{} and \logg{} values. The choice 
of \teff{} is the most important parameter
affecting the [O/Fe] ratios and the difference between the two oxygen
indicators. The [m/H] value is slightly significant, but it is unlikely 
that a 0.3 dex systematic error in the [m/H] value would occur in practice. 

The effects of various parameter changes on the forbidden and permitted
oxygen and Fe~II abundances as a function of \teff, \logg{}, and 
[Fe/H] are given in Plots 6--8 for each star in the sample. 

The most important feature in the plots is that a \teff{} change has 
opposite effects on the permitted and forbidden line abundances, while 
the other parameter changes affect the two indicators in similar ways. 
The figures also show that a systematic error in the surface gravity affects
the abundance indicators by the same amount, but a systematic \teff{} error 
affects giants more than dwarfs, and metal-poor stars more than metal-rich 
stars. An error in [m/H] affects metal-rich stars more than metal-poor ones. 
Fe~II mostly behaves like [O~I] when [m/H] is changes, but more like O~I when
\teff{} is altered. 
These results indicate that great care must be taken
when comparing the results from different evolutionary status. Systematic
parameter problems affect some stars, such as metal-poor giants, more
than others.

\subsubsection{Random Errors in [O/H] and [O/Fe]}

While systematic effects may be the most important factors in resolving
the the disagreement between the forbidden and permitted lines,
it is important to know the random errors associated with the O abundances
to determine the significance of discrepancies. We consider 
random errors in EWs, \teff, \logg{},
and [m/H] of the model. The abundance error due to $v_t$ is less 
than 0.03 dex and will be not considered in our analysis. We modify the
formula from \citet{m95}:

{\setlength
\arraycolsep{2pt}
\begin{eqnarray}
\sigma^2_{log\epsilon}= \sigma^2_{EW} +
\left({\partial log \epsilon\over\partial T}\right)^2 
\sigma^2_T + 
\left({\partial log\epsilon\over\partial \mlogg}\right)^2 
\sigma^2_{\mlogg}  + 
\nonumber\\
\left({\partial log\epsilon\over\partial \mfeh}\right)^2 \sigma^2_{\mfeh} + {} 
{} 2\biggl[\left({\partial log\epsilon\over\partial T}\right)
\left({\partial log\epsilon
\over\partial \mlogg}\right)\sigma_{T\mlogg} + 
\nonumber\\
 \left({\partial log\epsilon\over\partial \mfeh}\right) \left({\partial log\epsilon\over\partial 
\mlogg}\right) \sigma_{\mlogg \mfeh} + {}
\nonumber\\
{} \left({\partial log\epsilon\over\partial \mfeh}\right)\left({\partial 
log\epsilon\over T}\right) \sigma_{T \mfeh} \biggr],
\end{eqnarray}}

where $\sigma_{T \mlogg}$, for example, is defined as 

\begin{equation}
\sigma_{T \mlogg}=\frac{1}{N}\sum_{i=1}^N \left(T_i - \overline{T}\right)
\left(\mlogg_{\it i} - \overline{\mlogg}\right).
\end{equation}

The partial derivatives were calculated in Section 6.2.2.
Equation 1 shows that \logg{} is dependent on
\teff. The extent to which our uncertainties in \teff{} and \logg{} are
correlated depends on the magnitude of the error in \mV{}. Therefore, for
each star in our sample, we did a Monte Carlo experiment where we allowed
\teff{} and \mV{} to vary based on  their errors, then calculated \logg{} using
Equation 1, then placed into Equation 5 to calculate $\sigma_{\rm T \mlogg}$.  
Other $\sigma$ values were determined in the same manner. 

The errors in ratios such as [[O~I]/Fe] can be appreciably smaller than 
the addition in quadrature of [O~I] error and Fe~II error, 
because they have similar
sensitivities to changes in atmospheric parameters. We used Equation
A20 from \citet{m95}, modified to include [m/H] errors,
to calculate abundance ratio errors. The error bars
in Figure 10 are calculated using this formula.  No errors were calculated
for the Sun.

\subsection{Results for Alonso and Houdashelt Scales}

The abundance results from the Alonso and Houdashelt parameter scales are
shown in Figures 9 and 10. In Figure 9, neither scale results in
both sets of oxygen lines giving the same abundance for all stars, although
the warmer Houdashelt scale does a better job. The unweighted mean 
value of \opof{} ($\equiv \log{\rm n(O_p)} - \log{\rm n(O_f)}$) for the 
Alonso scale is $+0.35 \pm 0.03$ (sdom), and $+0.09 \pm 0.04$ (sdom) for the
Houdashelt scale.  In Figure 10, \opof{} is shown as a function of the 
stellar parameters.

The ratio \opof{} is larger for the cooler, lower
surface gravity giants than for the warmer, higher gravity subgiants and  
dwarfs.  Both least-squares and Spearman rank-order tests confirm that
there are highly significant anti-correlations between \opof{} and these two 
parameters.  These same tests do not support a correlation between \opof{} and
[Fe/H].  This is in contrast to previous studies (see Figure 1) in which the  
value of \opof{} increases with decreasing [Fe/H].  In these earlier studies, 
the forbidden oxygen abundances came from giants and the permitted 
abundances came from dwarfs.  Our result suggests that the growth of \opof{}
with decreasing [Fe/H] is at least partially due to comparing stars of 
different evolutionary status.

In Figure 10, it appears that the \opof{} distribution for the Houdashelt 
scale is bimodal, with
some stars clustered at \opof{} $= +0.5$ and the majority around \opof{} $= 0.0$.
Indeed, a KMM test \citep{ash} finds that there is a 96\% probability that
two Gaussians fit the distribution better than a single Gaussian.  The best 
fit model would place 12 stars in a group with a mean of $+0.51 \pm 0.12$, and 
the remaining 43 in a group with a mean of $-0.02 \pm 0.17$.  While this is 
only a two-sigma result, understanding why the 12 stars (all with \opof{} 
$> +0.3$) are outliers may yield clues to the origin of the overall problem.

Unfortunately, a detailed investigation into the properties these 12 stars
found nothing striking about these stars except that they all have 
\logg{} $< 3$ and [Fe/H] $< -1$.  These 12 stars are also among the stars with 
the highest \opof{} values when the Alonso \teff{} scale is applied, so the 
origin of the high \opof{} value may be unrelated to the Houdashelt scale.  
Checks for binarity, evolutionary status, systematic errors with the 
photometry, EW measurements,
\mV{} and reddening determinations, etc. did not yield any noticeable pattern
for the 12 stars, especially one that would lead to such a tight clustering
of outliers.  

It can be concluded here that both the Alonso and Houdashelt scales fail
to totally resolve the discrepancy.  While the warmer Houdashelt scale comes 
closer
than the Alonso scale, there are still several giant stars that have
large \opof{} values.  Two possible reasons for the failure are:  
First, there is missing input physics in the analysis and an additional 
correction to the abundance results is necessary, or, second, the physics of 
the analysis is adequate, but the input parameters for the models are 
incorrect.  

Full exploration of the first option is beyond the scope of this 
paper, but one simple explanation is that the NLTE corrections adopted
here are simply wrong.  To correct the differences seen in Figure 10, 
NLTE corrections would have to be much larger for low-gravity stars.  The
corrections of \citet{g00} are nearly identical to those of \citet{nlte} 
(see Figure 4) for this type of star, so the choice of NLTE correction does
not to affect the results.  In section 6.4, we will look at the effect of 
changing our choice of stellar atmospheres.
In Section 7, we will assume the second option is correct and 
calculate stellar parameters that reconcile the indicators.

\subsection{MARCS vs. Kurucz Atmospheres}

As mentioned above, the analyses to this point have been done using
Kurucz atmospheres. The MARCS grid of stellar atmospheres \citep{b76} 
are an independent calculation of one-dimensional, plane-parallel atmospheres. 
To test whether the adopted atmosphere grid makes a significant
difference, we re-analyzed the measured EW values through atmospheres
using the dereddened Alonso temperature scale parameters. 

The comparison between the results from Kurucz and MARCS models is shown in 
Figure 11.  The abundances derived from the permitted and forbidden oxygen 
and Fe~II lines are all slightly larger for the Kurucz models than for the 
MARCS models.  These tendencies are enhanced at lower metallicities.

The MARCS-derived oxygen abundances show a similar discrepancy
between the permitted and forbidden lines on average ($0.33 \pm 0.03$ for
MARCS compared to $0.35 \pm 0.03$ for the Kurucz models). Therefore, 
the use of MARCS models instead of Kurucz models will not
solve the problem. However, the MARCS models show a lower discrepancy 
for metal-poor giant stars, while the Kurucz model results show lower 
discrepancies for more metal-rich, less-evolved stars. If we used
the most favorable atmospheric model for a given 
star, the difference between the oxygen abundance indicators could be
reduced by up to $\sim 0.1$ dex in some cases.  However, there is no
justification for such a selective use of atmospheres.

\section{Ad Hoc Parameter Solution}

\subsection{Calculating the Parameters}

In Section 6.2.2 we analyzed the behavior of the derived oxygen abundances
as a function of the various stellar parameters.  We can use that knowledge
to derive an ad hoc parameter scale that forces the two oxygen indicators
to agree.  We will then examine the resulting parameters for their validity.

To derive the parameters, we assume that the changes in the abundances
with respect to parameter changes are all linear--that is, the first
partial derivatives are all constant. Therefore we can use the values
from Section 6.2.2 in the calculation.

If we define $X = \log{\rm n(species)}$ then

\begin{eqnarray}
\Delta X = \frac{\partial{X}}{\partial{T}} \Delta T_{\rm eff} + \frac{\partial{X}}{\partial{\log{g}}} \Delta \log{g} + 
\nonumber \\
\frac{\partial{X}}{\partial{\rm [m/H]}} \Delta {\rm [m/H]} + \frac{\partial{X}}{\partial{v_t}} \Delta v_t
\end{eqnarray}

From Table 8, it is clear that the difference in the
oxygen abundance indicators is most sensitive to the \teff{} value.
Therefore we will cast the above equation as a function of
\teff{} and the partials derived in Section 6.2.2.

Because it was found that the variation in oxygen abundances due to 
changes in $v_t$ is small, we will assume 
$\frac{\partial{X}}{\partial{v_t}} = 0$ and drop that term.
We will also assume that the dependency of \mV{} and the
bolometric correction on \teff{} is small and adopt
$\Delta \log{g} = \frac{4}{\ln{10}} \frac{\Delta \mteff}{\mteff}$
(derived from Equation 1).
The adopted atmospheric [m/H] value is just the [Fe~II/H] value. That
value changes like any other abundance, so it is itself described
by Equation 6, but in this case $\Delta X_{\rm [Fe/H]} = \Delta {\rm [m/H]}$. 
If that substitution is made, then it is possible to solve for 
$\Delta {\rm [m/H]}$ as a function of \dT. 
When these substitutions are made into Equation 6, we have 
an equation that describes the
change in the abundance of element X as only a function of \dT.
Therefore, we can solve for the
value of \teff{} which will force an agreement between the abundances
derived for the forbidden and permitted oxygen lines.

\subsection{Ad Hoc Scale Abundance Results}

The value of \opof{} was reduced to less than 0.01 dex in one to three
iterations of the above procedure.  
Table 11 gives the total value of \dT{} and the final stellar parameters,
while Table 12 gives the resulting abundances. 
The mean value of \dT{} derived from the Alonso scale results 
is $+213 \pm 134$ K (s.d.); Figure 12 plots the \dT{} value as a 
function of other parameters.  There is a
trend of increasing \dT{} with decreasing \logg{}, which is expected    
because the giants have the largest \opof{} values.

When the same method is applied using the Houdashelt \teff{} scale results,
the final \teff{} values are the same as the values calculated for 
the Alonso scale results.  The
final mean difference between the ``Ad Hoc'' scale and the Houdashelt
scale is $+58 \pm 168$ K (s.d.).  If the stars are split into the two groups
based on the Houdashelt results discussed in Section 6.3, the 12 stars 
with Houdashelt-scale \opof{} values of $> +0.3$ need their Houdashelt
\teff{} values increased, on mean, by $+305 \pm 69$ K (s.d.), while the
remaining 43 stars require a mean change of $-12 \pm 113$ K (s.d.).  
The resulting [O/Fe] vs. [Fe/H] plot is shown in Figure 13.

The calculation of the random errors given in Tables 11 and 12 and shown in 
Figure 12 and 13 required changes to the method used in Section 6.2.
We determined the random error in \teff{} by considering the uncertainties 
in the oxygen equivalent widths and in \mV{}.  However, when
calculating the errors for the O$_p$ and O$_f$ abundances, we needed 
to add three
terms to Equation 4 that took into account the correlation between the
error in the oxygen equivalent widths on one hand, and \teff{}, \logg{} and
[m/H] on the other.

\subsection{Comparision with Previous Results}

A comparison between our oxygen abundances for all three temperature scales
and those of several earlier
works is given in Figure 14.  With a few exceptions, the points lie on
parallel tracks to the 45-degree line.  
The effect of changes in the temperature, for example, can be seen by
comparing the top, middle, and lower panels.
As the temperature increases, the forbidden line oxygen 
abundances shift to higher abundances, while the permitted line abundances 
shift to lower abundances.

\citet{k93} proposed a temperature scale that would resolve the
discrepancy between the forbidden and permitted lines.  His calibration
of \teff$-$color relationships is valid for stars with $\mteff > 5470$ K,
so there are only 5 stars within our sample that can be compared to
the \citet{k93} scale.  The differences, T$_{\rm Ad Hoc} -$ T$_{\rm King}$,
are $+352$~K, $+260$~K, $+180$~K, $+165$~K, and $-140$~K for a mean
difference of $163 \pm 185$~K.  
King provides specific \teff{} values for three more stars in common 
with this sample.
If all eight stars are considered, the Ad Hoc scale is $+147 \pm 142$~K 
warmer, and there is only a weak significance to the correlation 
between the two scales.  While the overlap in samples is small, the
evidence suggests that the King and Ad Hoc temperature scales are not
in general agreement, even though both scales agree that increased \teff{}
values can resolve the oxygen problem.

\section{Does the Ad Hoc \teff{} scale make sense?}

The ad hoc temperature scale was picked to solve the oxygen problem,
but we must examine whether the scale reasonable when compared to 
other observations or the predictions of stellar evolution.

\subsection{\teff--\logg{} Plane}

In Figure 15, we plot the \logg{} vs. \teff{} plane for all three
parameter scales. Also plotted are 10 and 12 Gyr $\alpha$-enhanced isochrones
from \citet{v00} for [Fe/H] $= -0.84$, $-1.54$, and $-2.31$.
This range spans the observed [Fe/H] range for most of the target stars;
thus, most of the stars should lie between the isochrones.  The mean [Fe/H]
values for the Alonso, Houdashelt, and Ad Hoc scales are $-1.50$, $-1.51$
and $-1.54$, respectively. 

Many of the stars in the warmer Ad Hoc scale lie outside the range defined
by the isochrones. This is not a metallicity effect, because the mean [Fe/H]
values of all three scales are similar.
Agreement could be re-established by increasing
the \logg{} values by $\sim +0.5$ dex, because this would have a small effect
on \opof{} (see Table 8). The largest change due to a gravity increase
would be the $\sim +0.2$ dex increase in [Fe/H], but the net affect to 
\opof, on average, would be less than 0.05 dex. However, an
increase in \logg{} of $\sim 0.5$ dex would imply that the adopted
M$_{\rm bol}$ values were too bright by $\sim 1.2$ magnitudes. An error of
this size would be noticeable in the comparison of isochrones to globular
cluster sequences.

\subsection{\teff{} Values from Balmer Profiles}

The strength and profiles of the Balmer lines are dominated by the Stark 
effect and, theoretically, are very good temperature indicators 
\citep{gray,b02}. This indicator is 
insensitive to errors in the reddening and surface gravity, and is a
reasonably independent source of \teff{} values. Recent works that
include stars studied here include \citet[HIP~57939]{b02}
\citet[HIP~57939 and HIP~104659]{z00}, and 
\citet[HIP~30668, HIP~49371, HIP~98532, and HIP~104659]{f94}
Including all measurements,
the mean value of T$_{\rm Balmer}$ - T$_{\rm Alonso}$ is $+1 \pm 45$ K (sdom), 
while the mean value of T$_{\rm Balmer}$ - T$_{\rm Ad Hoc}$ is $-146 \pm 47$ K 
(sdom).  Again, like the Fe~I NLTE test above, the comparison stars are mostly
dwarfs and subgiants, so a more extensive study of Balmer line-based \teff{}
values would be welcome.

\subsection{Extra Reddening?}

If the reddening estimates assumed in Section 5.1 were too low, the resulting
\teff{} values would be too cool.  If the Ad Hoc temperature scale was
the correct one for the stars, then the \teff$-$color relations could be
inverted to give the intrinsic colors of the star.  The reddening could 
then be determined by comparison with the observed colors.

When this is done with the Alonso calibrations, we find that the required
mean increase in E(B$-$V) needed to account for the temperature change 
ranges from 0.05 to 0.12 mag, depending on the color used (greatest for
B$-$V, smallest for V$-$K).  The mean measured value of E(B$-$V) for the sample
is $0.04 \pm 0.06$, so the additional reddening required overall is 
larger than the original value.  

The star-to-star scatter in the values is large.  The star that requires
the largest increase in reddening is BD~+30~2611 (=HIP 73960), which
was found to need 0.32 mag of additional reddening, but the measured 
E(B$-$V) = 0.00.  Similarly, the closest sample star, HD~ 103095 (=HIP~57939),
also with measured E(B$-$V) = 0.00, would require 0.12 mag of additional 
reddening.  For the 8 sample stars with the most reliable Hipparcos parallaxes 
($\sigma_{\pi}/\pi < 0.10$, with a
mean distance of 42 pc), the mean additional reddening is 0.07 $\pm 0.05$ mag, 
while the mean measured reddening was 0.01 mag (five have measured
E(B$-$V) values of 0.00).  The mean additional reddening necessary for the
19 giant stars with \mV{} $< 0$ (mean distance of $\sim$ 750 pc) is 
0.19 $\pm$ 0.08 mag, while the mean measured E(B$-$V) is 0.04.

The overall increase of the reddening value, especially for nearby stars
that should not be heavily reddened, strongly indicates that additional 
reddening is not the source of the temperature difference.  Reddening for
individual stars can be very uncertain, and may be the cause for some of the
random scatter, but it is unlikely it is the cause of the systematic 
difference.  Studies of giants within globular clusters, for which the
distance and reddening can be better determined than for individual field
stars, could be used to help settle this issue.

\subsection{Summary of Comparisons}

For all the tests attempted here, the Alonso scale produced a better match to
the observations than the warmer Ad Hoc scale.  The Houdashelt scale lies 
between the other two. Therefore, it is hard to 
justify a major change in stellar parameters just to improve the oxygen 
abundance situation. If the Alonso parameters are the correct ones
to adopt, then we have to accept that a 1-D, LTE analysis with the 
presently available NLTE
corrections adopted here is not sufficient to analyze oxygen abundances.  

\section{Discussion}

Recently, \citet{n02} and \citet{i01} found
that same oxygen abundance was derived using either 
the permitted and forbidden lines in dwarfs and subgiants. They used 
analyses similar to our first attempt to derive abundances, i.e, they 
calculated temperatures from photometry, \logg{} from Equation 1, etc.
Our analysis of giants and subgiants shows that the two abundance
indicators have not been reconciled for all stars. As indicated by 
Figure 10 the greatest values for \opof{} are for the low-gravity, 
low-temperature giants.  For both the Alonso and Houdashelt
scales, we find \opof{} $\sim 0$ for the parameter space
explored by Nissen \etal{} and Israelian \etal{} (6000~$\pm$~100K,
\logg{} $\sim$ 4.0 dex and [Fe/H] $< -2.4$). 

However, there are some outstanding problems remaining even with the subdwarf 
and subgiant analyses. Kurucz and MARCS models do not give the same answers.
Nissen \etal{} calculated [O/Fe] $= 0.43$ for HD~189558 (= HIP~98532) using 
OSMARCS models.
Despite using similar atmospheric parameters and equivalent widths, we found
[O/Fe] $= 0.22$ when we used Kurucz model atmospheres. \citet{i01}
derived a smaller difference between the permitted O lines and forbidden lines
in BD~+23~3130 (HIP~85855) than we derive here, mainly because of the different
electron densities in the different sets of Kurucz models used. As mentioned
in the introduction, \citet{cg00} ascribed their difficulties
with cool giants in part to their use of the \citet{kz92} models.
\citet{n02} showed that 
the use of 3-D model atmospheres could alter the oxygen and iron
abundances in metal-poor dwarfs. The correction to [O$_{\rm f}$/Fe] 
for the metal-poor dwarf HD~140283 was $-0.26$ dex, and it is no longer clear 
whether the permitted and forbidden lines would produce the same oxygen 
abundance. 

Three-dimensional model atmospheres are not yet available for giants, and the
one-dimensional models, especially for the coolest giants, do not result in the
agreement seen in the higher gravity stars. We have discussed above why one
possible solution, a higher temperature scale, is not a good one.
\citet{l91} and \citet{nlte} suggest that [O~I] is filled in by
emission, and this option is discussed and eliminated in the Appendix.

There are still several solutions that can solve this discrepancy. 
Giants have large convection zones, so
granulation may play an even larger role than in the dwarfs. 
Giants have thin atmospheres that are penetrable by UV radiation, so NLTE
corrections for the permitted lines are important.  One-dimensional NLTE
calculations may not work if a three-dimensional model is needed to describe
the true conditions in the atmosphere.  

Until three-dimensional models become widely available, there are
still tests that can be done using traditional methods.  For example,
the gravities of giant stars have larger uncertainties than dwarfs due to
the uncertain distance to the stars.  \citet{k00} discusses whether, like 
dwarfs, the LTE Fe~I/Fe~II ionization balance
can no longer be used to derive surface gravities in metal-poor giants. 
As seen in Table 7, Hipparcos parallaxes are of little use to individual 
giants.  Although the changing the surface gravity is not the solution to 
resolving the \opof{} controversy, \logg{} is crucial for calculating the 
absolute O abundance. Thus, until more reliable data is
available from GAIA or SIM, a study of permitted vs. forbidden lines
in cluster stars with accurately known distances would be helpful.  
The chemical homogeneity of most clusters also  makes it possible to use the 
abundances other heavy elements to help constrain the parameters.

\section{Summary}
We have analyzed the forbidden and permitted oxygen lines in 55 stars, 
including dwarfs and giants and spanning [Fe/H] values from solar to $-2.7$
in an attempt to understand the discrepancy in these oxygen abundance
indicators.  We first tried a standard analysis using
the temperature scales of Alonso and Houdashelt.  These models produced
$<$[Op/Of]$>$ values of $+0.35$ and $+0.09$, respectively. 
The discrepancy was 
largest for cool giants, but evolved stars of all types
favor high \opof{} values. The \opof{} ratio is most sensitive to
temperature of all the atmospheric parameters, and it is the 
only one where the the effect of a change in the parameter is opposite 
for the two indicators. 

Using our understanding of the effects of parameter changes on the abundances,
we calculated a new parameter scale that would bring the two sets of oxygen 
lines into agreement.  These parameters, however, disagree with other 
temperature diagnostics, such as colors, the fits to the Balmer lines, 
and the bolometric luminosities. We 
conclude that either improved NLTE corrections for the permitted lines or
other phenomena, perhaps associated with convection and 
granulation, are needed to solve the oxygen problem.

\acknowledgements

JPF and JAJ would like to thank the staffs at the Canada-France-Hawaii 
Telescope, Mauna Kea Observatories, Lick Observatory, and
Kitt Peak for their invaluable assistance with the observations for this
project.  We also like to thank Poul Nissen and Garak Israelian for their
insightful correspondences on BD~+23~3130 and Robert Kraft, Chris Sneden,
and James Hesser for their comments on drafts of this paper.  Finally, we
gladly thank the anonymous referee for his valuable and insightful comments.
This research has made use of the SIMBAD database, operated at CDS, 
Strasbourg, France.  This publication makes 
use of data products from the Two Micron All Sky Survey, which is a joint 
project of the University of Massachusetts and the Infrared Processing and 
Analysis Center, funded by the National Aeronautics and Space Administration 
and the National Science Foundation.

\appendix

\section{Emission in the [O~I] Features?}

\citet{l91} suggested that emission from circumstellar shells could
fill in the [O~I] lines in giant stars.  These shells are the result of 
mass loss on the giant branch.  The resulting lower EW values would
then lead to the discrepancy in the oxygen abundance indicators. 

The \citet{l91} model proposes that a mass loss rate of a few $10^{-7}$ 
\msun{} yr$^{-1}$ could create an H~I region of about 32 AU around the
giant.  If the temperature of this region was about the same as the giant
(4500 K in the model) and the density is about $6.8 \times 10^6$ cm$^{-3}$,
the amount of photons emitted by the 6300 \AA{} [O~I] line from the H~I
region would reduce the measured EW by 20 m\AA.
\citet{l91} admits that the required mass loss rate is a 
factor of about five too high than expected by theory, but the remaining 
assumptions are not wildly unreasonable.

We therefore examined the 6300.31 \AA{} region of the 16 stars with 
M$_{\rm V} < 0$ observed with Gecko for signs of emission. 
The stellar absorption lines of our sample are resolved at the spectral 
resolution of Gecko.  For example, in the 16 giants 
examined here, the [O~I] 6300.31 \AA{} absorption lines have a mean FWHM of 
$0.175 \pm 0.026$ \AA{} ($\sim 10$ pixels).  The telluric [O~I] emission lines 
in these same spectra have a mean FWHM of $0.0625 \pm 0.002$ \AA{} ($\sim 3$ 
pixels).  

The dominant line-broadening mechanism in the H~I region is thermal Doppler 
broadening, which for this case would be 0.045 \AA, or less than the 
instrumental profile of Gecko.  Therefore, any emission from an H~I region 
surrounding the giant should be a narrow feature.  
Regions of 1 \AA{} ($\sim 55$ pixels), centered at 6300.31 \AA{} for these 16 
giants are shown in Figures 16 and 17.  No binning or smoothing has been
applied to the spectra.  As can be seen, no significant emission 
is present.  

Finally, if the 6300.31 \AA{} [O~I] line was producing significant emission,
other emission lines may be present.  \citet{l91} estimates that the 
chromospheric H$_{\alpha}$ emission (which dominates over the H$_{\alpha}$ 
emission from the H~I region) from the model system would be several 
\AA ngstroms in equivalent width.  Therefore, we examined the H$_{\alpha}$
lines of the 16 giants in the lower resolution spectra used to measure the
Fe and permitted O~I lines.  Of these giants, only six show any sign of having
asymmetric H$_{\alpha}$ profiles (HIP~17639 is not among these six stars). 
Of these six, only two, BD~+30~2611 (= HIP~73960) and HD~165195 (= HIP~88527) 
show any sign of H$_{\alpha}$ emission.  For these two giants, the 
6300.31 \AA{} [O~I] profile is deep, symmetric, and free of obvious emission. 
We therefore conclude that emission from an H~I region as described by 
\citet{l91} does not affect the equivalent width of the 
[O~I] lines to any significant amount.


\clearpage
\begin{deluxetable}{rrrccccccccl}
\tabletypesize{\tiny}
\tablenum{1}
\tablewidth{0pt}
\tablecaption{Observation Log}
\tablehead{
\colhead{HIP}&\colhead{HD}&\colhead{BD}&\multicolumn{4}{c}{[O I] Data} & \multicolumn{4}{c}{O I and Fe Data} & \colhead{Add.} \\
 &  & &\colhead{Instr.\tablenotemark{a}} &\colhead{Date} &\colhead{Exp} & \colhead{S/N\tablenotemark{b}}  &\colhead{Instr.\tablenotemark{a}} &\colhead{Date} &\colhead{Exp} & \colhead{S/N} & \colhead{Data} \\
 &  & & &\colhead{UT} &\colhead{(s)} & \colhead{Per Pixel}  & &\colhead{UT} &\colhead{(s)} & \colhead{Per Pixel} & \colhead{Refs.} }
\startdata
Sun\tablenotemark{c} & \nodata &\nodata & KPNO &2002 Jan 5 & 1800 &400 & same &\nodata & \nodata & 375 & \\ 
434 & 20 &\nodata & Gecko &2001 Oct 1 & 1800 &145 & KPNO &2002 Jan 6 & 1800 & 100 & 1  \\ 
484 & 97 &-20 6718 & Gecko &2001 Oct 1 & 1800 &145 & KPNO &2002 Jan 7 & 1800 & 100 & \\ 
2413 & 2665 &+56 70 & Gecko &2001 Sep 29 & 900 &250 & F00 &\nodata & \nodata & 150 & 1 \\ 
2463 & 2796 &-17 70 & Gecko &2001 Sep 29 & 900 &250 & Ham. &2000 Aug 12 & 900 & 120 & \\ 
        &         &        & \nodata &\nodata & \nodata &\nodata & KPNO &2002 Jan 8 & 900 & 200 & \\ 
3985 & 4906 &+18 111 & Gecko &2001 Sep 30 & 900 &225 & Ham. &2000 Aug 12 & 1800 & 110 & \\ 
4933 & 6268 &$-28$ 322 & Gecko &2001 Sep 30 & 1800 &250 & KPNO &2002 Jan 8 & 600 & 200 & \\
\enddata
\tablenotetext{a}{``KPNO'' designates data from the KPNO 4-m and echelle 
spectrograph, ``Gecko'' from CFHT and the Gecko spectrograph,  
``Ham.'' from the Lick 3-m and Hamilton spectrograph, and ``same''
means the same spectra was used for all the measurements.  Many of the 
Hamiliton spectra were first observed for Fulbright (2000; F00) and
Johnson (2002; J02).  See text for more details.}
\tablenotetext{b}{The S/N measure for the Gecko data only is the mearsure of all exposures combined into a single spectrum.}
\tablenotetext{c}{The solar spectrum as reflected off the asteroid Vesta.}
\tablenotetext{d}{Permitted O I line EW measurments come exclusively from literture sources.}
\tablerefs{(1) Gratton et al. 2000, (2) Israelian et al. 2001, (3) Takeda et al. 2000, (4) Cavallo et al. 1997, (5) Tomkin et al. 1992, (6) Cayrel et al. 2001.}
\end{deluxetable}

\clearpage
\begin{deluxetable}{rrrrrrrr}
\tabletypesize{\tiny}
\tablenum{2}
\tablewidth{0pt}
\tablecaption{Oxygen Line Equivalent Widths}
\tablehead{
\colhead{HIP}&\colhead{6300 \AA}&\colhead{6363 \AA}& \colhead{$\sigma_{\rm [O I]}$} & \colhead{7772 \AA} & \colhead{7774 \AA} & \colhead{7775 \AA} & \colhead{$\sigma_{\rm O I}$} \\
 &\colhead{m\AA}&\colhead{m\AA}& \colhead{m\AA} & \colhead{m\AA} & \colhead{m\AA} & \colhead{m\AA} & \colhead{m\AA} }
\startdata
Sun & 4.5 & \nodata & 0.6 & 70.8 & 61.0 & 47.2 & 1.1 \\ 
434 & 4.9 & \nodata & 1.3 & 44.9 & 46.4 & 47.1 & 4.0 \\ 
484 & 10.1 & 4.2 & 1.3 & 26.4 & 17.3 & 13.7 & 3.1 \\ 
2413 & 3.4 & \nodata & 0.8 & 7.7 & 6.5 & 4.4 & 2.0 \\ 
2463 & 6.0 & 2.3 & 0.8 & 12.4 & \nodata & \nodata & 2.0 \\ 
3985 & 5.9 & 2.7 & 0.9 & 40.7 & 32.6 & 20.9 & 2.7 \\ 
4933 & 5.1 & \nodata & 0.8 & 7.2 & 6.3 & \nodata & 2.0 \\ 
5445 & 5.6 & 1.7 & 0.8 & 17.3 & 14.3 & 9.3 & 1.9 \\
\enddata
\end{deluxetable}

\clearpage 
\begin{deluxetable}{llrrrrrrrrrrrrrr} 
\tablenum{3a} 
\tabletypesize{\tiny} 
\tablewidth{0pt} 
\tablecaption{Equivalent Widths (m\AA)} 
\tablehead{ 
& &  \multicolumn{1}{c}{SUN/}  & \multicolumn{1}{c}{HIP}    & \multicolumn{1}{c}{HIP}    & \multicolumn{1}{c}{HIP}    & \multicolumn{1}{c}{HIP}    & \multicolumn{1}{c}{HIP}    & \multicolumn{1}{c}{HIP}    & \multicolumn{1}{c}{HIP}    & 
 \multicolumn{1}{c}{HIP}  & \multicolumn{1}{c}{HIP}    \nl 
  & & \multicolumn{1}{c}{VESTA} & \multicolumn{1}{c}{   434} & \multicolumn{1}{c}{   484} & \multicolumn{1}{c}{  2463} & \multicolumn{1}{c}{  3985} & \multicolumn{1}{c}{  4993} & \multicolumn{1}{c}{  5445} & \multicolumn{1}{c}{  5458} & 
\multicolumn{1}{c}{ 17649} & \multicolumn{1}{c}{ 43228}  } 
\startdata 
 FeI  & 4531.15 & \nodata &  90 & \nodata &  79 & \nodata &  85 & \nodata & \nodata & \nodata & \nodata \nl 
 FeI  & 4592.66 & \nodata & \nodata & \nodata &  65 & \nodata &  69 & \nodata & \nodata & \nodata & \nodata \nl 
 FeI  & 4595.36 &  38 & \nodata &  50 & 8.9 & \nodata & \nodata &  24 & \nodata &  22 & \nodata \nl 
 FeI  & 4602.01 &  71 &  32 &  68 &  19 &  68 &  19 &  41 & \nodata &  72 &  75 \nl 
 FeI  & 4602.94 & \nodata & \nodata & \nodata &  74 & \nodata &  89 & \nodata & \nodata & \nodata & \nodata \nl 
\enddata 
\tablenotetext{a}{[The complete version of this table is in the electronic 
edition of the Journal.  The printed edition contains only a sample.]}
\end{deluxetable}

\clearpage 
\begin{deluxetable}{llrrrrrrrrrrrrrr} 
\tablenum{3b} 
\label{tab:table3b}
\tabletypesize{\tiny}
\tablewidth{0pt} 
\tablecaption{Equivalent Widths (m\AA)} 
\tablehead{ 
& &  \multicolumn{1}{c}{HIP}  & \multicolumn{1}{c}{HIP}    & \multicolumn{1}{c}{HIP}    & \multicolumn{1}{c}{HIP}    & \multicolumn{1}{c}{HIP}    & \multicolumn{1}{c}{HIP}    & \multicolumn{1}{c}{HIP}    & \multicolumn{1}{c}{HIP}    & 
 \multicolumn{1}{c}{HIP}  & \multicolumn{1}{c}{HIP}    \nl 
  & & \multicolumn{1}{c}{ 58514} & \multicolumn{1}{c}{ 62235} & \multicolumn{1}{c}{ 65852} & \multicolumn{1}{c}{ 74491} & \multicolumn{1}{c}{ 88977} & \multicolumn{1}{c}{ 91182} & \multicolumn{1}{c}{ 94931} & \multicolumn{1}{c}{ 96248} & 
\multicolumn{1}{c}{106095} & \multicolumn{1}{c}{107337}  } 
\startdata 
 FeI  & 4531.15 & \nodata & \nodata & \nodata & \nodata & \nodata & \nodata & \nodata & \nodata & \nodata & \nodata \nl 
 FeI  & 4592.66 & \nodata & \nodata & \nodata & \nodata & \nodata & \nodata & \nodata &  62 & \nodata & \nodata \nl 
 FeI  & 4595.36 & \nodata &  24 & \nodata & \nodata &  36 & \nodata &  59 & \nodata & \nodata & \nodata \nl 
 FeI  & 4602.01 &  73 &  35 &  57 & \nodata &  51 &  89 &  72 & \nodata &  66 & \nodata \nl 
 FeI  & 4602.94 & \nodata & \nodata & \nodata & \nodata & \nodata & \nodata & \nodata &  75 & \nodata & \nodata \nl 
\enddata 
\tablenotetext{a}{[The complete version of this table is in the electronic 
edition of the Journal.  The printed edition contains only a sample.]}
\end{deluxetable} 

\clearpage
\begin{deluxetable}{rrrrrrrrrrr}
\tabletypesize{\tiny}
\tablenum{4}
\tablewidth{0pt}
\tablecaption{Photometric Values}
\tablehead{
\colhead{HIP}&\colhead{V$_0$}&\colhead{E(B-V)}&\colhead{(B-V)$_0$}&\colhead{(b-y)$_0$}&\colhead{(V-R)$_0^J$}&\colhead{(V-R)$_0^C$}&\colhead{(V-I)$_0^J$}&\colhead{(V-I)$_0^C$}&\colhead{(V-K)$_0$}&\colhead{c$_0$} }
\startdata
434 & 9.04 & 0.01 & 0.68 & 0.427 &  \nodata  & \nodata & 0.73 & 0.94 & 1.79 & 0.473 \\
484 & 9.66 & 0.00 & 0.79 & 0.513 & 0.68 & 0.47 & 0.82 & 1.05 & 2.16 & 0.349 \\
2413 & 7.73 & 0.07 & 0.68 & 0.497 & 0.67 & 0.46 & 0.67 & 0.85 & 2.07 & 0.350 \\
2463 & 8.49 & 0.03 & 0.68 & 0.520 & 0.69 & 0.48 & 0.72 & 0.93 & 2.14 & 0.490 \\
3985 & 8.76 & 0.04 & 0.74 & 0.453 &  \nodata  & \nodata & 0.76 & 0.97 & \nodata & 0.291 \\
4933 & 8.09 & 0.02 & 0.80 & 0.583 &  \nodata  & \nodata & 0.81 & 1.04 & \nodata & 0.514 \\
5445 & 7.72 & 0.03 & 0.67 & 0.464 & 0.59 & 0.40 & 0.70 & 0.90 & 1.96 & 0.302 \\
5458 & 6.75 &  \nodata  & 1.14 & 0.735 & 0.95 & 0.66 & 1.12 & 1.44 & 2.88 & 0.487 \\
\enddata
\end{deluxetable}

\clearpage
\begin{deluxetable}{rccccccccrrrr}
\tabletypesize{\tiny}
\tablenum{5}
\tablewidth{0pt}
\tablecaption{Alonso Scale Parameters}
\tablehead{
\colhead{HIP}&\colhead{[m/H]$_{init}$}&\colhead{T$_{\rm B-V}$}&\colhead{T$_{\rm b-y}$}&\colhead{T$_{\rm V-R}$}&\colhead{T$_{\rm V-I}$}&\colhead{T$_{\rm V-K}$}&\colhead{T$_{\rm avg}$}&\colhead{T$_{\rm eff}$}&\colhead{$\sigma_{\rm T}$}&\colhead{B.C.}&\colhead{log $g$}&\colhead{v$_t$} }
\startdata
434 &$-1.4$& 5089 & 5334 & \nodata & 5426 & 5366 & 5304 & 5375 &  75 & $-0.25$& 2.41 & 1.90 \\
484 &$-1.2$& 4922 & 4983 & 4907 & 5172 & 4914 & 4980 & 4950 &  75 &$-0.35$& 2.68 & 1.40 \\
2413 &$-1.9$& 5042 & 4800 & 4769 & 5639 & 5024 & 5055 & 5025 & 75  &$-0.35$& 2.12 & 1.45 \\
2463 &$-2.2$& 4996 & 4833 & 4806 & 5462 & 4944 & 5008 & 4950 & 90  &$-0.37$& 2.13 & 1.80 \\
3985 &$-0.8$& 5143 & 5293 & \nodata & 5449 & \nodata & 5295 & 5225 & 106 &$-0.27$& 3.66 & 0.80 \\
4933 &$-2.1$& 4791 & 4671 & \nodata & 5207 & \nodata & 4890 & 4750 & 155 &$-0.44$& 1.42 & 2.10 \\
5445 &$-1.5$& 5109 & 5080 & 5145 & 5526 & 5150 & 5202 & 5175 & 75  &$-0.30$& 2.88 & 1.50 \\
5458 &$-1.0$& 4385 & 4326 & 4270 & 4487 & 4282 & 4350 & 4300 &  81 &$-0.62$& 1.63 & 1.55 \\
\enddata
\end{deluxetable}

\clearpage
\begin{deluxetable}{rcccccccrrrr}
\tabletypesize{\tiny}
\tablenum{6}
\tablewidth{0pt}
\tablecaption{Houdashelt Scale Parameters}
\tablehead{
\colhead{HIP}&\colhead{[m/H]$_{init}$}&\colhead{T$_{\rm B-V}$}&\colhead{T$_{\rm V-R}$}&\colhead{T$_{\rm V-I}$}&\colhead{T$_{\rm V-K}$}&\colhead{T$_{\rm avg}$}&\colhead{T$_{\rm eff}$}&\colhead{$\sigma_{\rm T}$}&\colhead{B.C.}&\colhead{log $g$}&\colhead{v$_t$} }
\startdata
434 &$-1.4$& 5590 & \nodata& 5506 & 5385 & 5494 & 5475 & 103 &$-0.24$& 2.45 & 1.90 \\
484 &$-1.2$& 5327 & 5053 & 5252 & 4955 & 5147  & 5125 & 172 &$-0.30$& 2.74 & 1.40 \\
2413 &$-1.9$& 5603 & 5087 & \nodata& 5057 & 5249  & 5075 &  75 &$-0.33$& 2.21 & 1.45 \\
2463 &$-2.2$& 5588 & 5016 & 5543 & 4979 & 5282  & 5225 & 329 &$-0.31$& 2.27 & 1.80 \\
3985 &$-0.8$& 5456 & \nodata& 5544 & \nodata & 5500 & 5500 &  75 &$-0.21$& 3.80 & 0.80 \\
4933 &$-2.1$& 5294 & \nodata& 5286 & \nodata & 5290 & 5300 &  75 &$-0.30$& 1.63 & 2.10 \\
5445 &$-1.5$& 5619 & 5403 & 5606 & 5180 & 5452  & 5300 & 158 &$-0.27$& 2.95 & 1.50 \\
5458 &$-1.0$& 4573 & 4327 & 4538 & 4321 & 4440  & 4425 & 134 &$-0.53$& 1.64 & 1.55 \\
\enddata
\end{deluxetable}

\clearpage
\begin{deluxetable}{rrrrrrrrrllr}
\tabletypesize{\tiny}
\tablenum{7}
\tablewidth{0pt}
\tablecaption{Absolute Magnitude}
\tablehead{
\colhead{HIP}&\colhead{$\pi$}&\colhead{$\frac{\sigma_{\pi}}{\pi}$}&\colhead{M$_V^{\pi}$}&\colhead{$\sigma$M$_V^{\pi}$}&\colhead{M$_V^{\rm H98}$}&\colhead{M$_V^{\rm ATT}$}&\colhead{M$_V^{\rm HB}$}&\colhead{M$_V^{\rm Iso}$}&\colhead{M$_V$}&\colhead{$\sigma$M$_V$}&\colhead{Mass} \\
 & \colhead{mas} & & & & & & & & & & \colhead{M$_{\odot}$} }
\startdata
434 & 1.40 & 0.94 &$-0.23$& 2.04 & +0.7 & +0.73 & 0.71 & \nodata & +0.7 & 0.2 & 0.60 \\
484 & 0.26 & 5.42 &$-3.27$& 11.77 & +1.4 & +1.28 & \nodata & +2.6 & +1.4 & 0.3 & 0.87 \\
2413 & 2.11 & 0.43 &$-0.65$& 0.93 & +0.2 & +0.66 & AGB? & +0.2 & +0.2 & 0.3 & 0.83 \\
2463 & 0.06 & 17.7 &$-7.62$& 38.37 &$-0.6$&$-0.81$& 0.56 & \nodata & +0.6 & 0.2 & 0.60 \\
3985 & 9.04 & 0.15 & +3.54 & 0.33 & \nodata & \nodata & \nodata & +3.6 & +3.6 & 0.3 & 0.97 \\
4933 & 1.92 & 0.54 &$-0.49$& 1.17 &$-0.9$&$-1.06$& AGB & \nodata &$-1.0$& 0.3 & 0.60 \\
5445 & 7.74 & 0.12 & +2.16 & 0.26 & +1.8 & 1.47 & \nodata & +1.7 & +1.8 & 0.3 & 0.84 \\
5458 & 4.35 & 0.18 &$-0.06$& 0.39 &$-0.6$& \nodata & \nodata &$-0.7$&$-0.5$& 0.3 & 0.89 \\
\enddata
\end{deluxetable}

\clearpage
\begin{deluxetable}{lrrrrrrrccc}
\tabletypesize{\tiny}
\tablenum{8}
\tablewidth{0pt}
\tablecaption{Mean Abundance Sensitivities}
\tablehead{
\colhead{Param. Change}&\colhead{[O I]}&\colhead{[O I]}&\colhead{O I}&\colhead{O I}&\colhead{O I}&\colhead{Fe I}&\colhead{Fe II}&\colhead{O$_f$/Fe II}&\colhead{O$_p$/Fe II}&\colhead{[O$_p$/O$_f$]} \\
 & \colhead{6300 \AA}&\colhead{6363 \AA}&\colhead{7772 \AA}&\colhead{7774 \AA}&\colhead{7775 \AA}&\colhead{Mean}&\colhead{Mean}& \colhead{Mean} & \colhead{Mean} & \colhead{Mean} }
\startdata
T$_{\rm eff} + 200$ K & +0.12 & +0.07 &$-0.22$&$-0.20$&$-0.15$& +0.23 &$-0.02$& +0.12  &$-0.17$&$-0.29$ \\
log $g$ + 0.3 dex   & +0.11 & +0.08 & +0.12 & +0.11 & +0.09 &$-0.01$& +0.12 &$-0.02$ &$-0.01$& +0.01  \\
{\rm [m/H]} + 0.3 dex     & +0.09 & +0.07 &  0.00 &$-0.01$&$-0.01$&$-0.01$& +0.05 & +0.03  &$-0.06$&$-0.09$ \\
v$_t$ + 0.3 km s$^{-1}$    &  0.00 &  0.00 &$-0.02$&$-0.01$&$-0.01$&$-0.06$&$-0.03$& +0.03  & +0.02 &$-0.01$ \\
\enddata
\end{deluxetable}

\clearpage
\begin{deluxetable}{rccccccccccccccc}
\tabletypesize{\tiny} 
\tablenum{9}
\tablewidth{0pt}
\tablecaption{Alonso Scale Results}
\tablehead{
\colhead{HIP}&\colhead{$\epsilon$(O$_f$)\tablenotemark{a}}&\colhead{$\sigma$}&\colhead{$\epsilon$(O$_p$)}&\colhead{$\sigma$}&\colhead{$\epsilon$(Fe I)}&\colhead{$\sigma$}&\colhead{$\epsilon$(Fe II)}&\colhead{$\sigma$}&\colhead{[Fe/H]}&\colhead{[O$_f$/Fe]}&\colhead{$\sigma$}&\colhead{[O$_p$/Fe]}&\colhead{$\sigma$}&\colhead{[O$_p$/O$_f$]}&\colhead{$\sigma$}}
\startdata
Sun & $8.83$ & \nodata & $8.79$ & \nodata & $7.54$ & \nodata &  $7.51$ & \nodata & $-0.01$ & $0.15$ & \nodata & $0.11$ & \nodata & $-0.04$ & \nodata \\
434 & $7.60$ & $0.10$ & $8.16$ & $0.15$ & $6.03$ & $0.07$ & $6.09$ & $0.04$ & $-1.43$ & $0.34$ & $0.07$ & $0.90$ & $0.17$ & $+0.56$ & $0.25$ \\
484 & $7.93$ & $0.12$ & $8.13$ & $0.07$ & $6.19$ & $0.08$ & $6.24$ & $0.07$ & $-1.28$ & $0.52$ & $0.06$ & $0.72$ & $0.10$ & $+0.19$ & $0.16$ \\
2413 & $7.09$ & $0.31$ & $7.26$ & $0.24$ & $5.62$ & $0.06$ & $5.52$ & $0.11$ & $-2.00$ & $0.40$ & $0.20$ & $0.57$ & $0.34$ & $+0.17$ & $0.54$ \\
2463 & $7.33$ & $0.19$ & $7.57$ & $0.31$ & $5.22$ & $0.10$ & $5.42$ & $0.05$ & $-2.10$ & $0.74$ & $0.14$ & $0.98$ & $0.36$ & $+0.24$ & $0.50$ \\
3985 & $8.37$ & $0.08$ & $8.60$ & $0.12$ & $6.93$ & $0.08$ & $6.75$ & $0.07$ & $-0.77$ & $0.45$ & $0.06$ & $0.68$ & $0.08$ & $+0.23$ & $0.14$ \\
4933 & $6.79$ & $0.30$ & $7.20$ & $0.09$ & $5.10$ & $0.12$ & $5.10$ & $0.14$ & $-2.42$ & $0.52$ & $0.16$ & $0.93$ & $0.21$ & $+0.41$ & $0.37$ \\
5445 & $7.73$ & $0.10$ & $7.79$ & $0.22$ & $6.00$ & $0.05$ & $5.93$ & $0.06$ & $-1.59$ & $0.63$ & $0.07$ & $0.69$ & $0.22$ & $+0.06$ & $0.29$ \\
\enddata
\tablenotetext{a}{$\epsilon({\rm X}) = \log{\rm n(X)} + 12$}
\end{deluxetable}

\clearpage
\begin{deluxetable}{rccccccccccccccc}
\tabletypesize{\tiny}
\tablenum{10}
\tablewidth{0pt}
\tablecaption{Houdashelt Scale Results}
\tablehead{
\colhead{HIP}&\colhead{$\epsilon$(O$_f$)}&\colhead{$\sigma$}&\colhead{$\epsilon$(O$_p$)}&\colhead{$\sigma$}&\colhead{$\epsilon$(Fe I)}&\colhead{$\sigma$}&\colhead{$\epsilon$(Fe II)}&\colhead{$\sigma$}&\colhead{[Fe/H]}&\colhead{[O$_f$/Fe]}&\colhead{$\sigma$}&\colhead{[O$_p$/Fe]}&\colhead{$\sigma$}&\colhead{[O$_p$/O$_f$]}&\colhead{$\sigma$}}
\startdata
Sun & $8.83$ & \nodata & $8.79$ & \nodata & $7.54$ & \nodata & $7.51$ & \nodata & $-0.01$ & $0.15$ & \nodata & $0.11$ & \nodata & $-0.04$ & \nodata \\
434 & $7.65$ & $0.10$ & $8.04$ & $0.15$ & $6.14$ & $0.08$ & $6.09$ & $0.04$ & $-1.43$ & $0.39$ & $0.07$ & $0.78$ & $0.17$ & $+0.39$ & $0.25$ \\
484 & $8.02$ & $0.12$ & $7.97$ & $0.07$ & $6.40$ & $0.11$ & $6.24$ & $0.07$ & $-1.28$ & $0.61$ & $0.06$ & $0.56$ & $0.10$ & $-0.05$ & $0.16$ \\
2413 & $7.13$ & $0.31$ & $7.22$ & $0.24$ & $5.67$ & $0.06$ & $5.53$ & $0.11$ & $-1.99$ & $0.43$ & $0.20$ & $0.52$ & $0.34$ & $+0.09$ & $0.54$ \\
2463 & $7.54$ & $0.19$ & $7.29$ & $0.31$ & $5.52$ & $0.15$ & $5.46$ & $0.05$ & $-2.06$ & $0.91$ & $0.14$ & $0.66$ & $0.36$ & $-0.25$ & $0.50$ \\
3985 & $8.46$ & $0.08$ & $8.32$ & $0.12$ & $7.22$ & $0.12$ & $6.69$ & $0.07$ & $-0.83$ & $0.60$ & $0.06$ & $0.46$ & $0.08$ & $-0.15$ & $0.14$ \\
4933 & $7.28$ & $0.30$ & $6.73$ & $0.09$ & $5.67$ & $0.22$ & $5.22$ & $0.14$ & $-2.30$ & $0.89$ & $0.16$ & $0.34$ & $0.21$ & $-0.55$ & $0.37$ \\
5445 & $7.78$ & $0.10$ & $7.65$ & $0.22$ & $6.13$ & $0.07$ & $5.92$ & $0.06$ & $-1.60$ & $0.69$ & $0.07$ & $0.56$ & $0.22$ & $-0.12$ & $0.29$ \\
\enddata
\end{deluxetable}

\clearpage
\begin{deluxetable}{rrrrrrr}
\tabletypesize{\tiny}
\tablenum{11}
\tablewidth{0pt}
\tablecaption{Ad Hoc Parameter Scale}
\tablehead{
\colhead{HIP}&\colhead{T$_{\rm AH}$}&\colhead{log $g$}&\colhead{[m/H]}&\colhead{v$_t$}&\colhead{$\Delta$T}&\colhead{$\sigma_{\Delta {\rm T}}$} }
\startdata
Sun & $5741$ & $4.4$ & $+0.0$ & $1.00$ & $ -34$ & \nodata \\
434 & $5728$ & $2.4$ & $-1.2$ & $1.90$ & $+353$ & 65 \\
484 & $5076$ & $2.6$ & $-1.1$ & $1.40$ & $+126$ & 60 \\
2413 & $5125$ & $2.2$ & $-1.9$ & $1.45$ & $+100$ & 90 \\
2463 & $5079$ & $2.2$ & $-2.0$ & $1.80$ & $+129$ & 60 \\
3985 & $5368$ & $3.7$ & $-0.6$ & $0.80$ & $+143$ & 70 \\
4933 & $4975$ & $1.4$ & $-2.3$ & $2.10$ & $+225$ & 70 \\
5445 & $5219$ & $2.9$ & $-1.4$ & $1.50$ & $ +44$ & 65 \\
\enddata
\end{deluxetable}

\clearpage
\begin{deluxetable}{rccccccccccccccc}
\tabletypesize{\tiny}
\tablenum{12}
\tablewidth{0pt}
\tablecaption{Ad Hoc Scale Results}
\tablehead{
\colhead{HIP}&\colhead{$\epsilon$(O$_f$)}&\colhead{$\sigma$}&\colhead{$\epsilon$(O$_p$)}&\colhead{$\sigma$}&\colhead{$\epsilon$(Fe I)}&\colhead{$\sigma$}&\colhead{$\epsilon$(Fe II)}&\colhead{$\sigma$}&\colhead{[Fe/H]}&\colhead{[O$_f$/Fe]}&\colhead{$\sigma$}&\colhead{[O$_p$/Fe]}&\colhead{$\sigma$}&\colhead{[O$_p$/O$_f$]}&\colhead{$\sigma$}}
\startdata
  Sun & $8.83$ & \nodata & $8.83$ & \nodata & $7.51$ & \nodata & $7.52$ & \nodata & $+0.00$ & $0.14$ & \nodata & $0.14$ & \nodata & $+0.00$ & \nodata \\
  434 & $7.83$ & $0.07$ & $7.82$ & $0.10$ & $6.37$ & $0.17$ & $6.14$ & $0.04$ & $-1.38$ & $0.52$ & $0.06$ & $0.51$ & $0.11$ & $-0.01$ & $0.12$ \\
  484 & $7.97$ & $0.08$ & $8.03$ & $0.08$ & $6.35$ & $0.19$ & $6.23$ & $0.06$ & $-1.29$ & $0.57$ & $0.01$ & $0.63$ & $0.11$ & $+0.05$ & $0.12$ \\
 2413 & $7.16$ & $0.08$ & $7.16$ & $0.14$ & $5.72$ & $0.23$ & $5.53$ & $0.05$ & $-1.99$ & $0.46$ & $0.10$ & $0.46$ & $0.15$ & $+0.00$ & $0.16$ \\
 2463 & $7.41$ & $0.06$ & $7.43$ & $0.13$ & $5.36$ & $0.18$ & $5.42$ & $0.03$ & $-2.10$ & $0.82$ & $0.07$ & $0.84$ & $0.13$ & $+0.02$ & $0.14$ \\
 3985 & $8.41$ & $0.09$ & $8.48$ & $0.10$ & $7.08$ & $0.20$ & $6.72$ & $0.07$ & $-0.80$ & $0.52$ & $0.11$ & $0.59$ & $0.12$ & $+0.07$ & $0.14$ \\
 4933 & $6.98$ & $0.09$ & $6.97$ & $0.11$ & $5.35$ & $0.19$ & $5.13$ & $0.06$ & $-2.39$ & $0.68$ & $0.11$ & $0.67$ & $0.12$ & $-0.01$ & $0.14$ \\
 5445 & $7.73$ & $0.09$ & $7.74$ & $0.12$ & $6.05$ & $0.18$ & $5.92$ & $0.06$ & $-1.60$ & $0.64$ & $0.11$ & $0.65$ & $0.13$ & $+0.01$ & $0.15$ \\
\enddata
\end{deluxetable}



\clearpage
\begin{figure}[bht]
\plotone{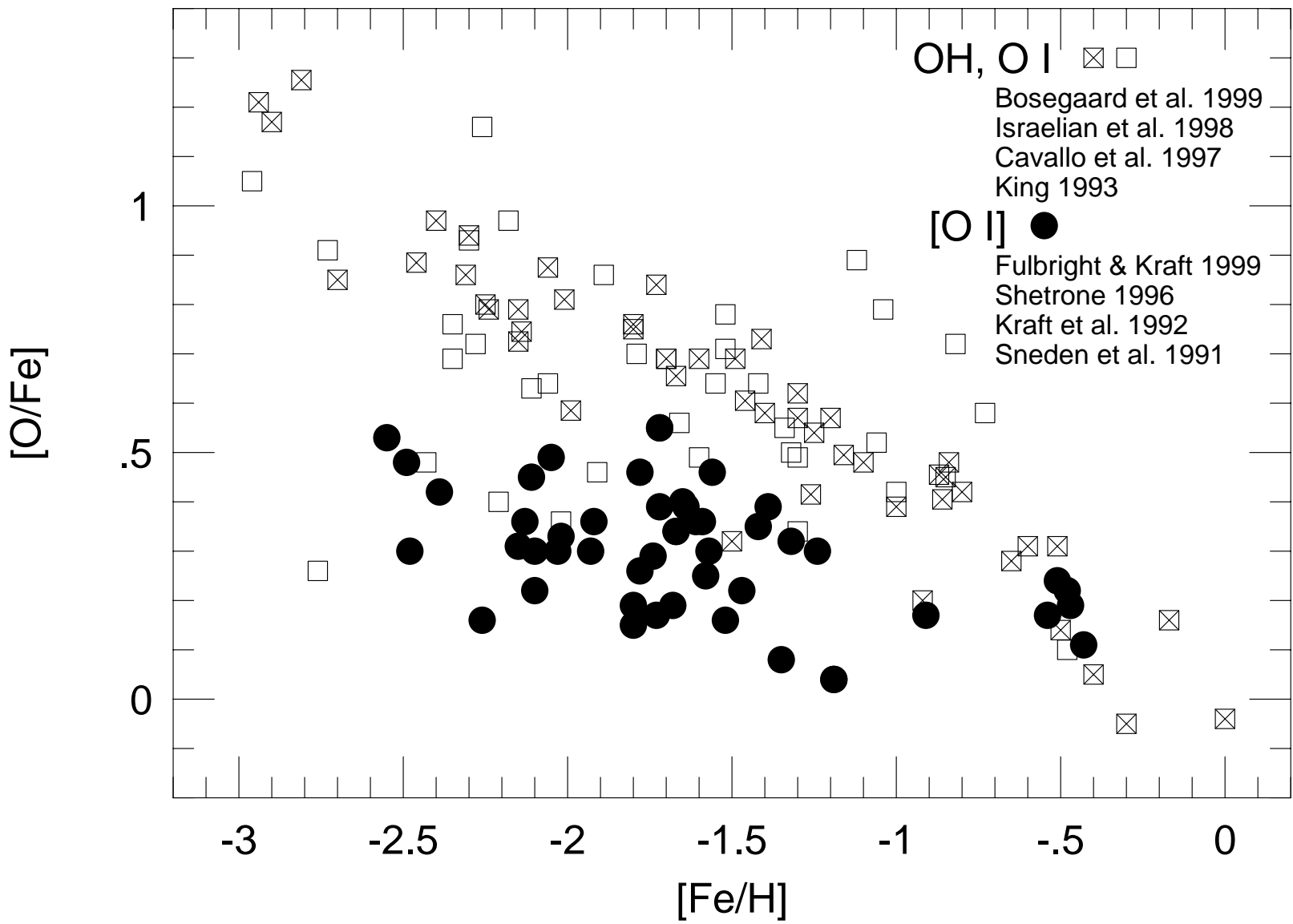}
\caption{Sample oxygen abundances derived by previous
studies to demostrate the systematic difference observed between the
forbidden [O I] lines and the permitted O I and molecular OH lines. }
\end{figure}

\clearpage
\begin{figure}[bht]
\plotone{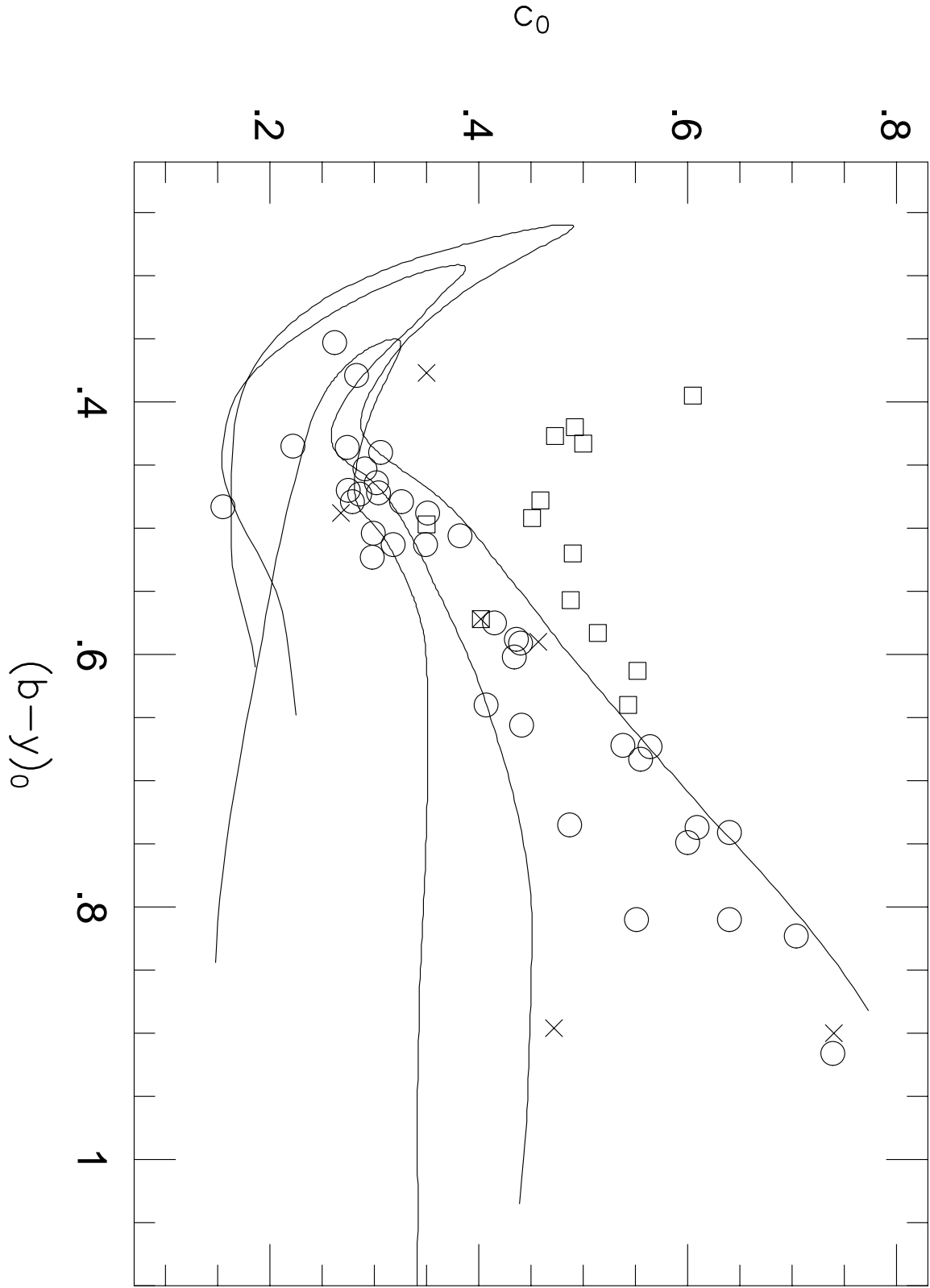}
\caption{Plot of the $c_0$ vs. $(b-y)_0$ values for the target 
stars. Candidate HB and AGB stars are plotted as squares. Other stars with 
reddening corrections are plotted with open circles, while those plotted as 
crosses have not been corrected for reddening, as no reddening data 
are available.  The lines are 12 Gyr $\alpha$-enhanced 
isochrones from
Clem \& Vandenberg (private communication) with [Fe/H] = $-0.71$, $-1.54$,
and $-2.31$.}
\end{figure}

\clearpage
\begin{figure}[bht]
\plotone{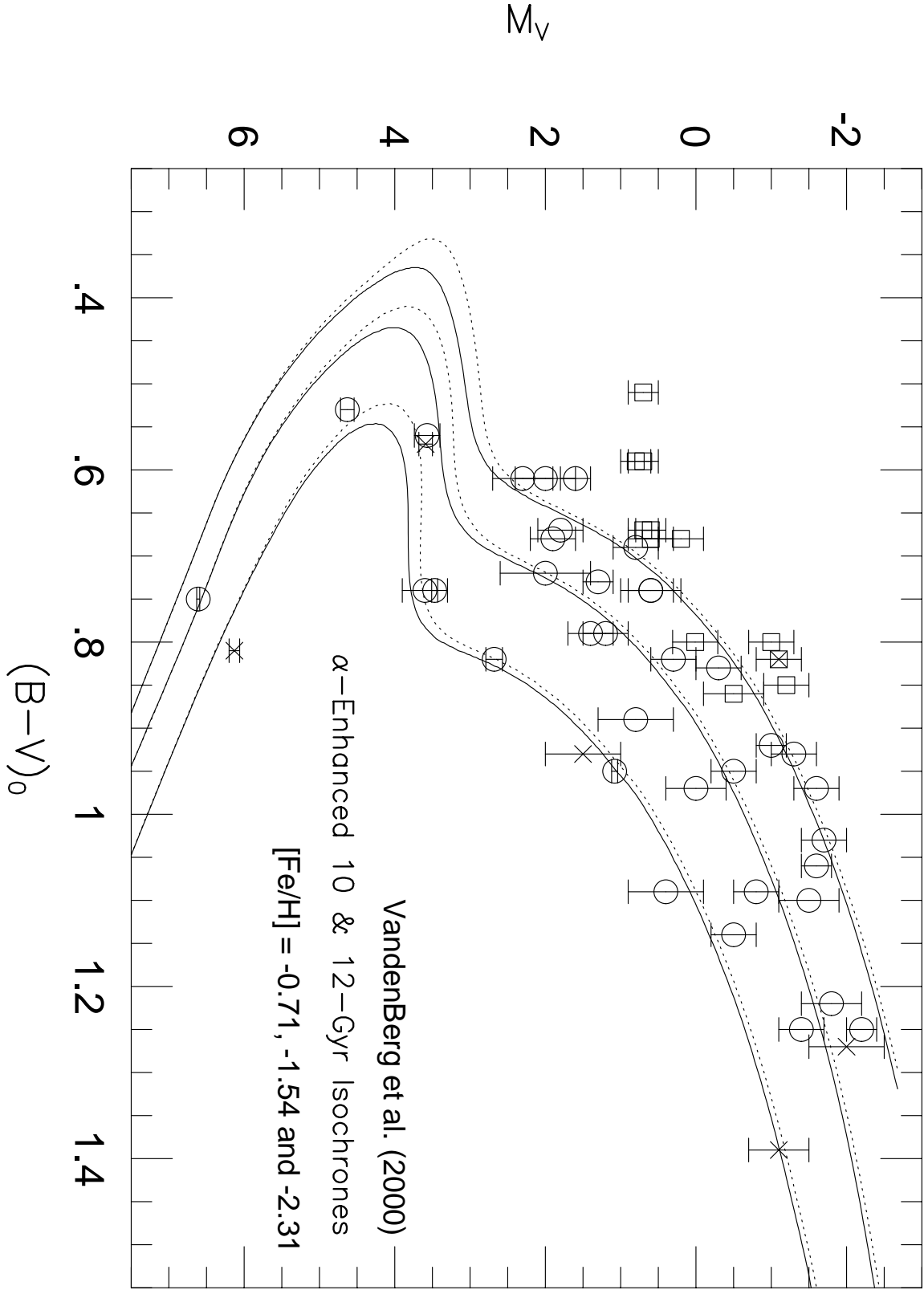}
\caption{Color--absolute magnitude diagram for the final
adopted values of the survey stars. Symbols are the same as in Figure 1.
The lines, from blue to red, represent the \citet{v00} 10- (dotted) and 12-Gyr
(solid) isochrones with [Fe/H] $= -2.31$, $-1.54$, and $-0.71$. }
\end{figure}

\clearpage
\begin{figure}[bht]
\plotone{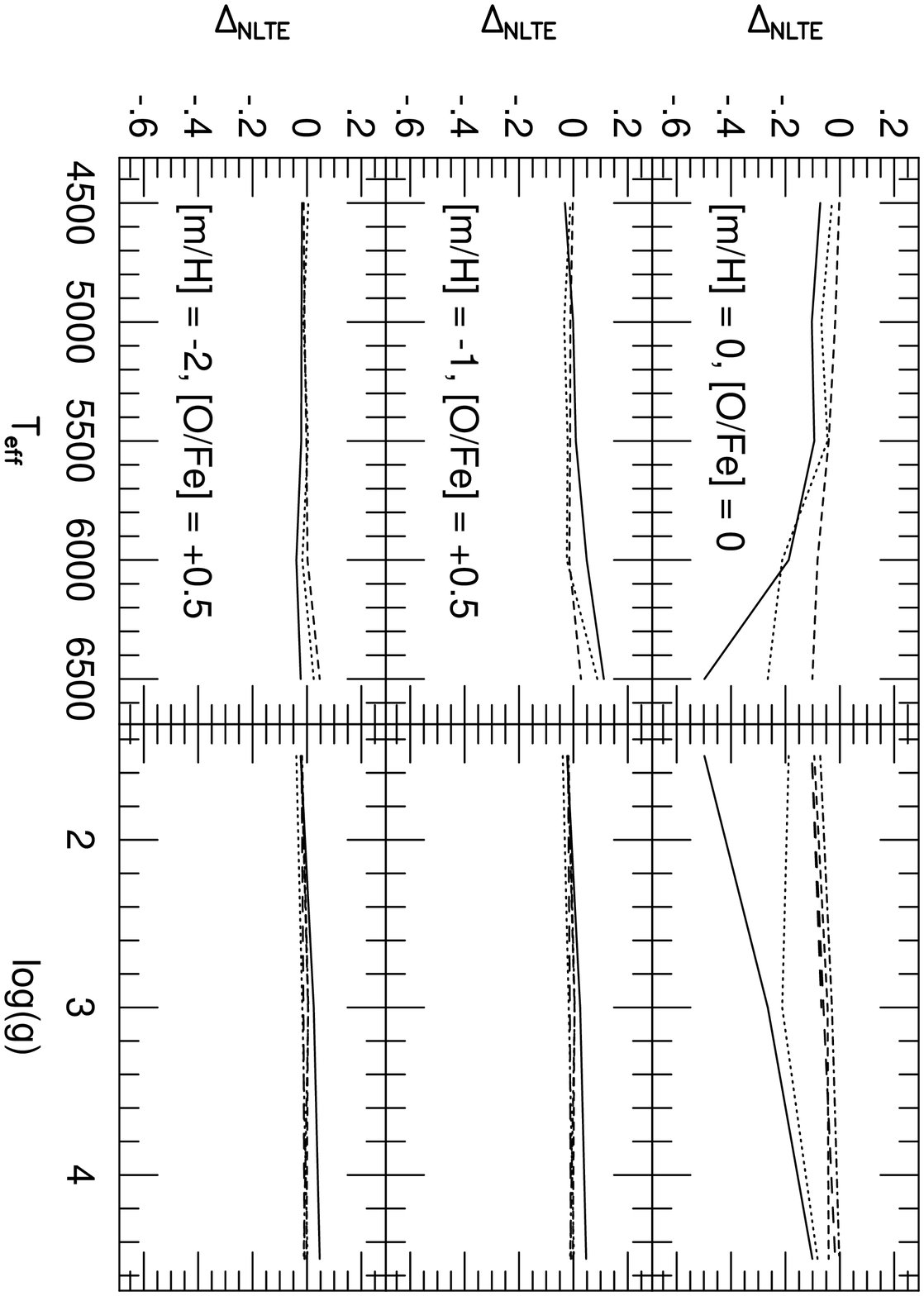}
\caption{The difference of the 
Takeda \etal{} NLTE correction minus the  
Gratton \etal{} correction, assuming the same O~I line strength  
for the 7772 \AA{} O~I line.
In the left column, the solid, dotted, and short-dash lines are 
for \logg{} = 1.5, 3.0, and 4.5, respectively. In the right column, the solid, 
dotted, short-dash, long-dash, and dotted-dash lines are for \teff{} $=$
6500 K, 6000 K, 5500 K, 5000 K, and 4500 K, respectively.
Except for hot low-gravity stars, the two corrections are similar. }
\end{figure}

\clearpage
\begin{figure}[bht]
\plotone{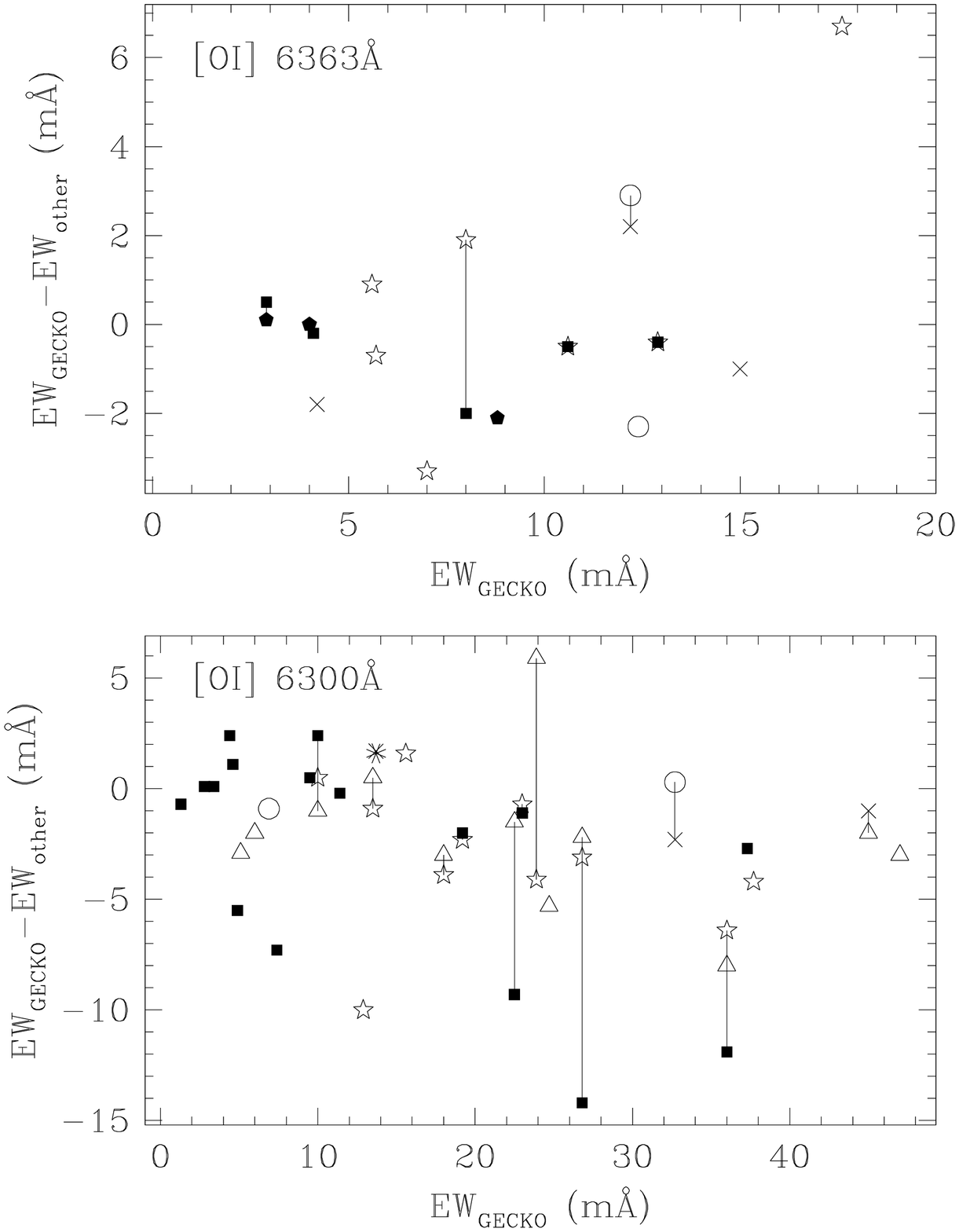}
\caption{EW comparison. The symbols are \citealt{g00}
(filled squares), \citealt{b88} (open triangles), \citealt{g86}
(asteriks), \citealt{s91} (open circles), \citealt{s96} (crosses),
\citealt{k92} (stars), and \citealt{nlte} (filled pentagons).  Solid lines
connect independent measurements of the same star.} 
\end{figure}

\clearpage
\begin{figure}[bht]
\plotone{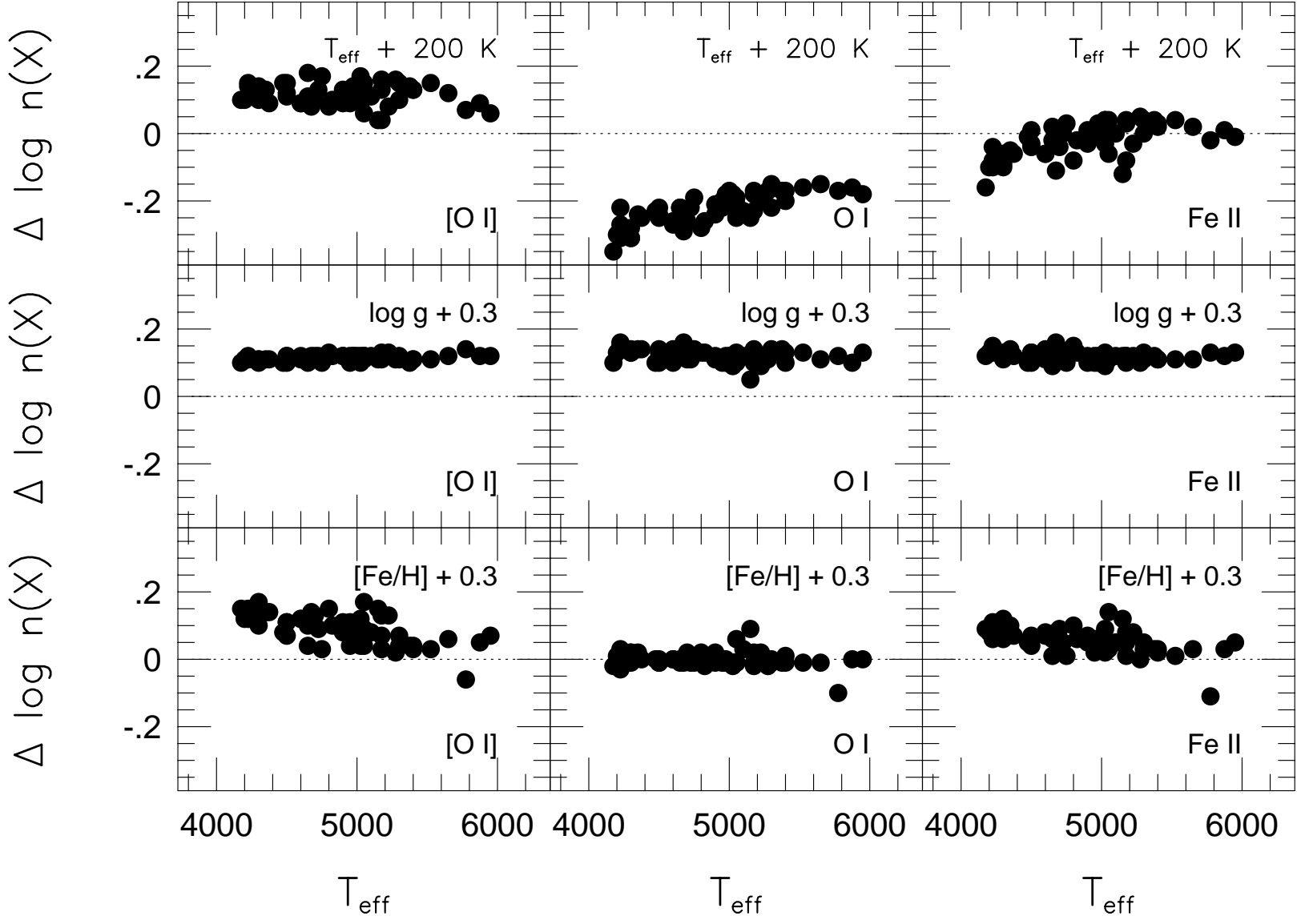}
\caption{The effect of a specific change in the stellar parameter
(each row) on the resulting abundance for each indicator (column) plotted as
a function of \teff. Each point is an individual star in this work.}
\end{figure}

\clearpage
\begin{figure}[bht]
\plotone{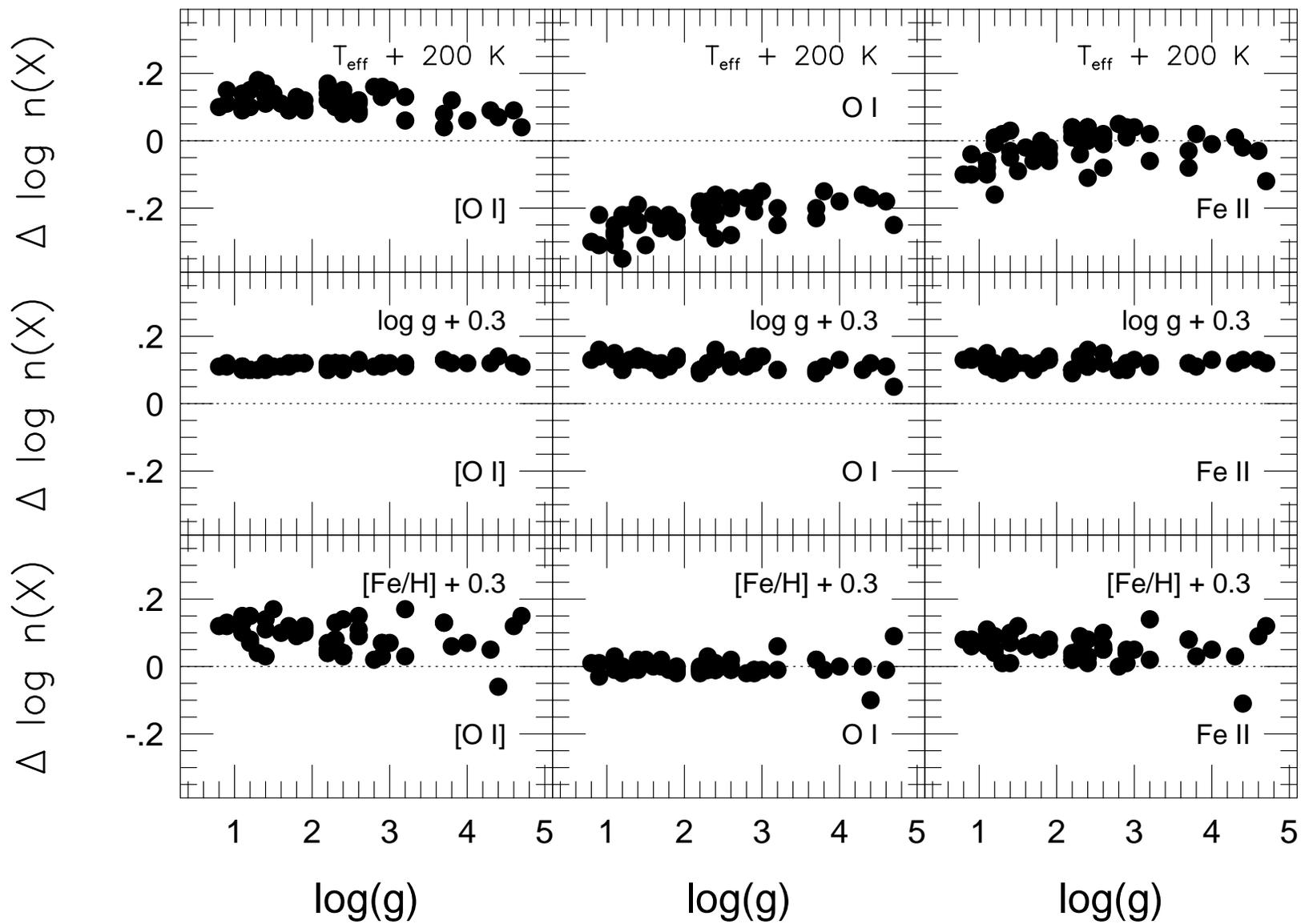}
\caption{Same as Figure 6, but plotted as a function of \logg{}.}
\end{figure}

\clearpage
\begin{figure}[bht]
\plotone{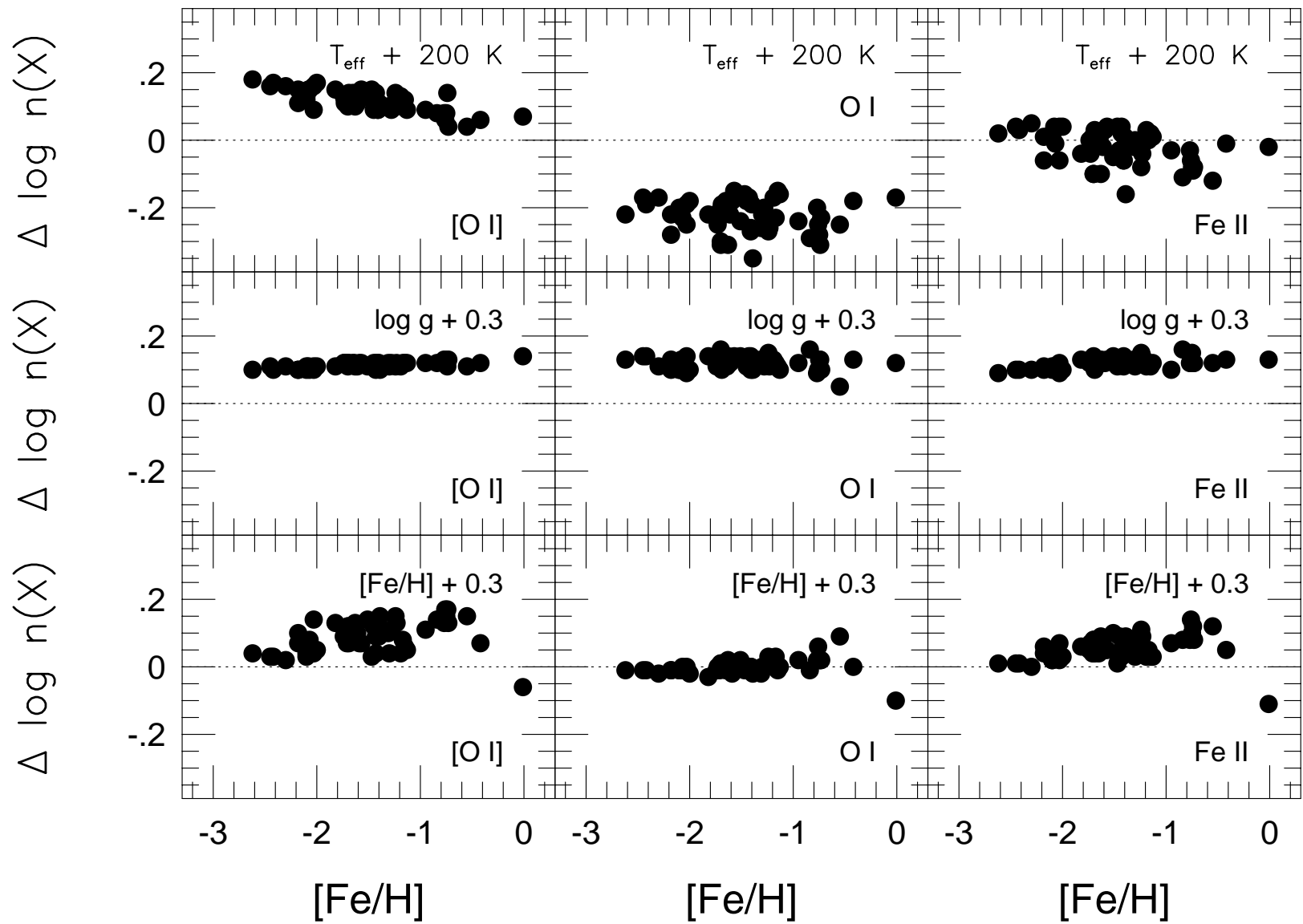}
\caption{Same as Figure 6, but plotted as a function of [Fe/H].}
\end{figure}

\clearpage
\begin{figure}[bht]
\plotone{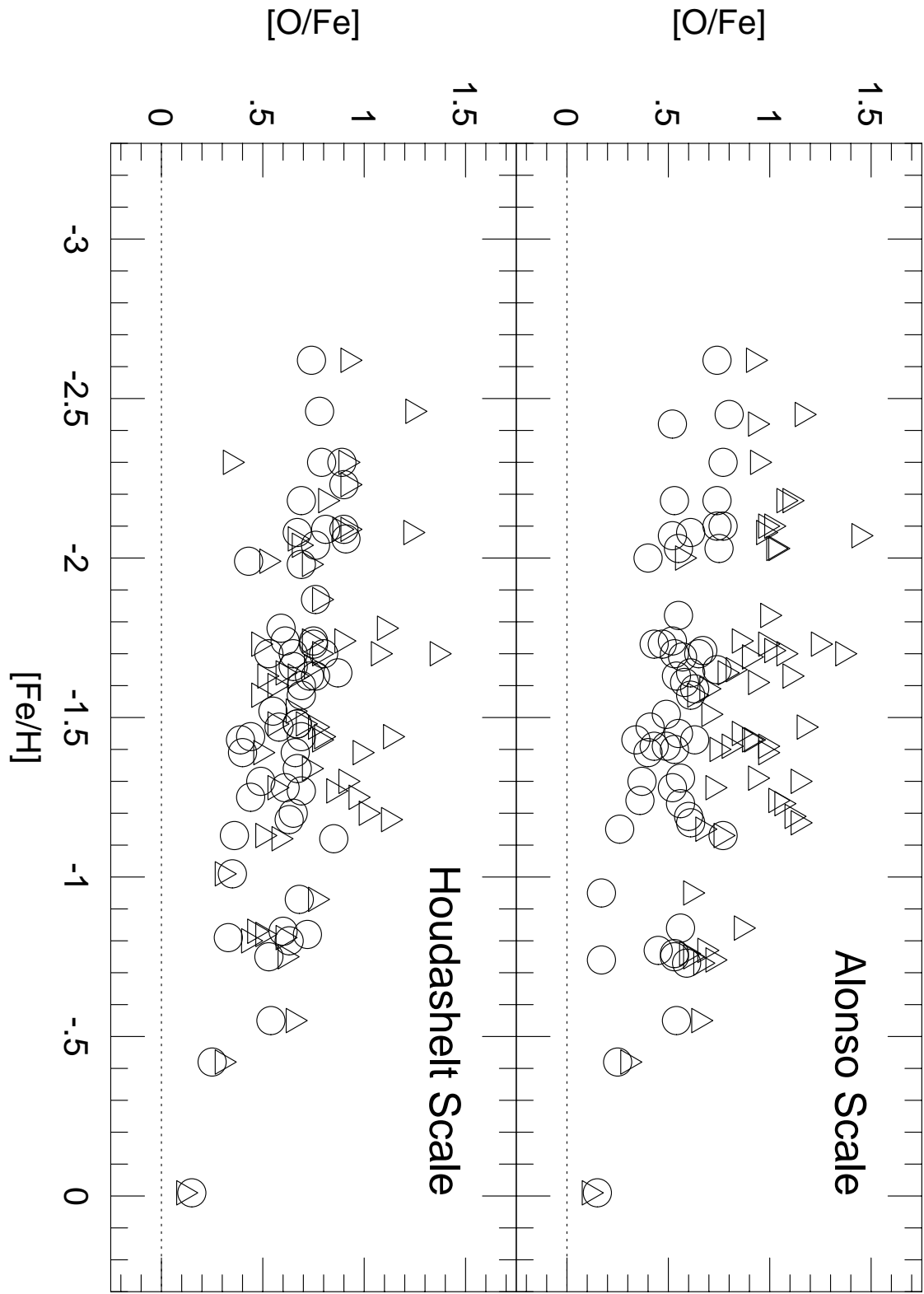}
\caption{Plot of the [O/Fe] vs. [Fe/H] values derived from
the permitted (triangles) and forbidden (circles) lines. Error bars have
been omitted for clarity. Both temperature scales show a difference in
the resulting oxygen abundnaces between the two indicators.}
\end{figure}

\clearpage
\begin{figure}[bht]
\plotone{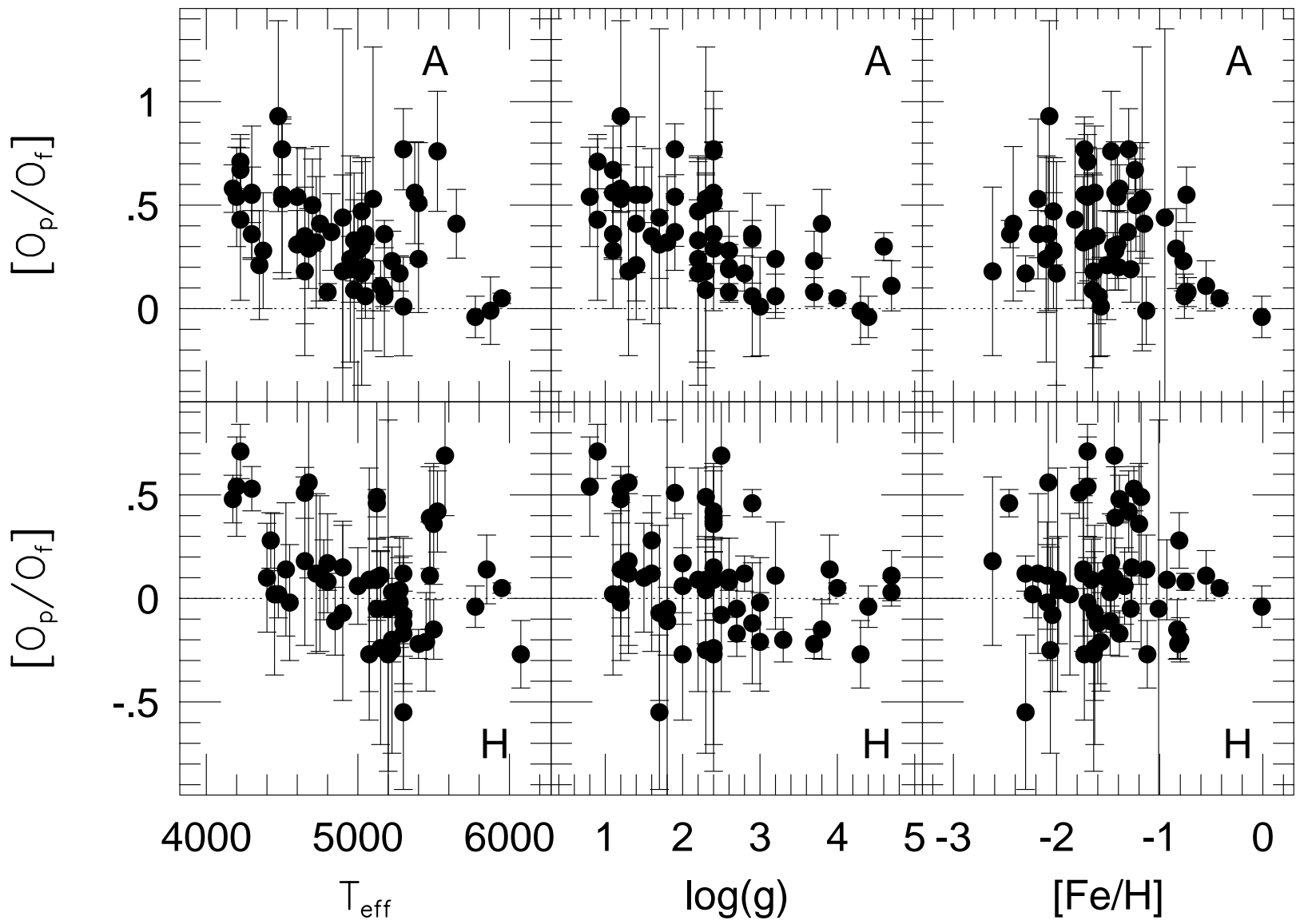}
\caption{The difference in the oxygen abundances as a function
of stellar parameters for the two temperature scales. }
\end{figure}

\clearpage
\begin{figure}[bht]
\plotone{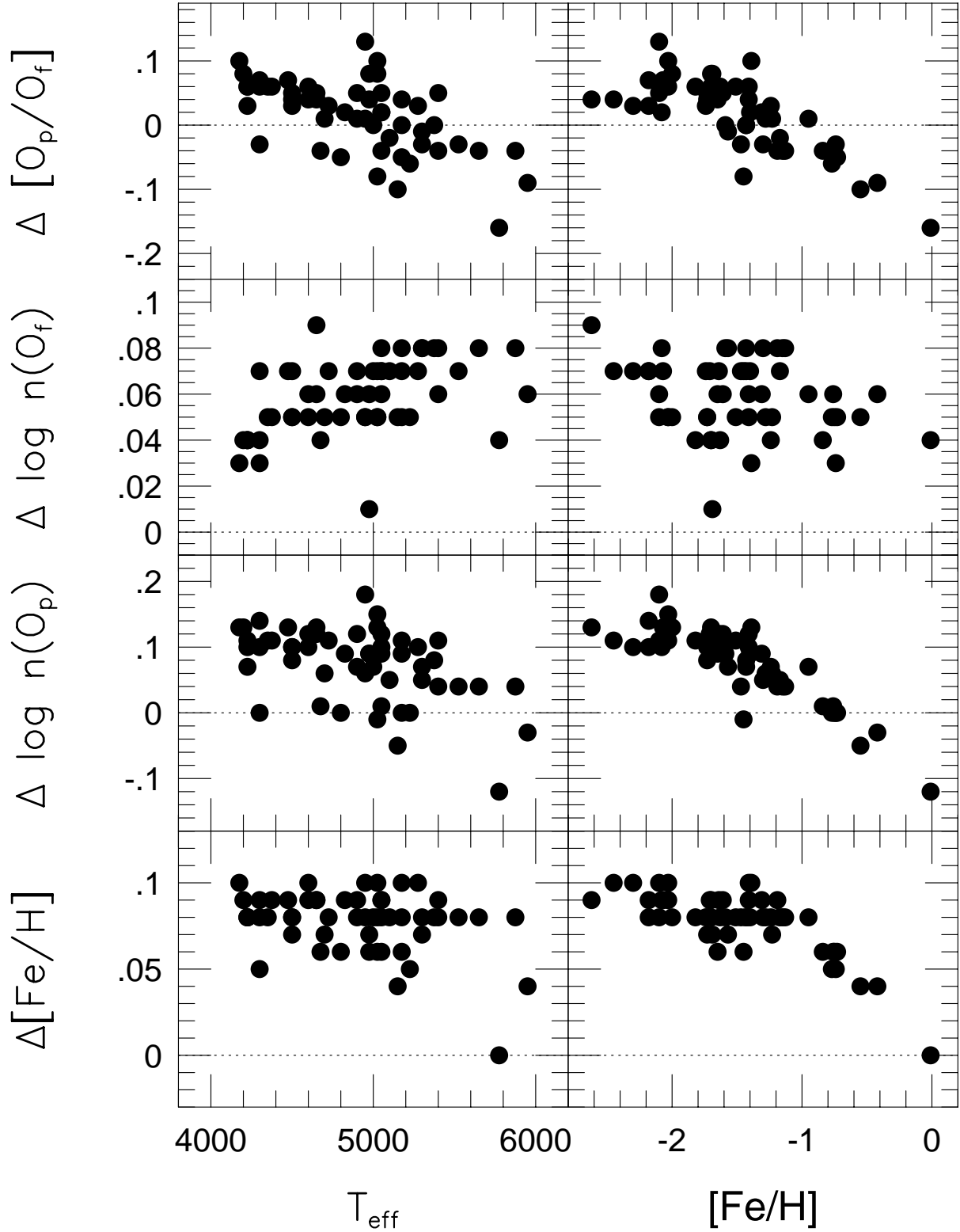}
\caption{The difference in the oxygen abundances derived
from the Kurucz and MARCS atmospheres (Kurucz minus MARCS).
Note the change of vertical scale for each row.}
\end{figure}

\clearpage
\begin{figure}[bht]
\plotone{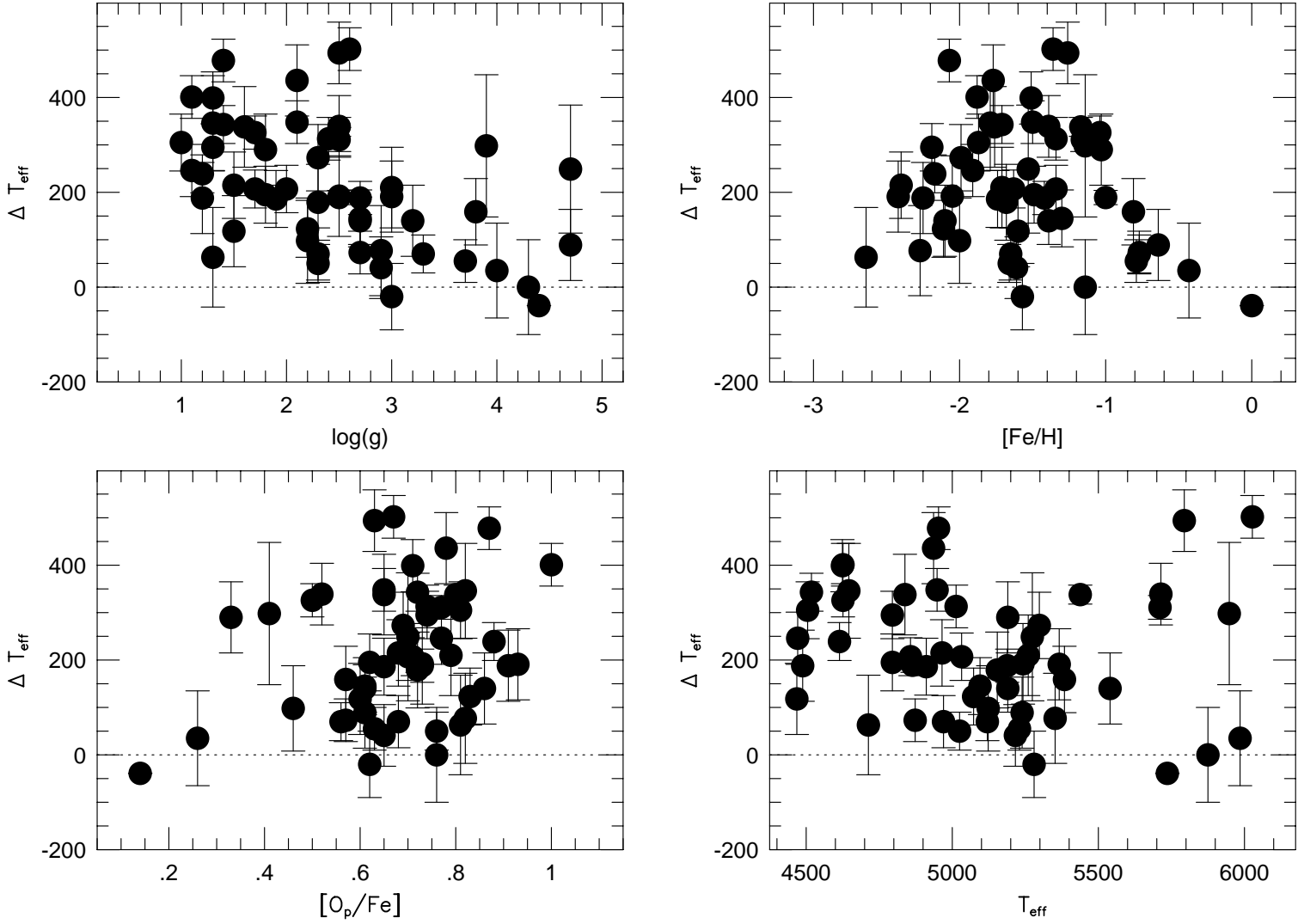}
\caption{Required change in the adopted \teff{} value
as a function of stellar parameters.} 
\end{figure}

\clearpage
\begin{figure}[bht]
\plotone{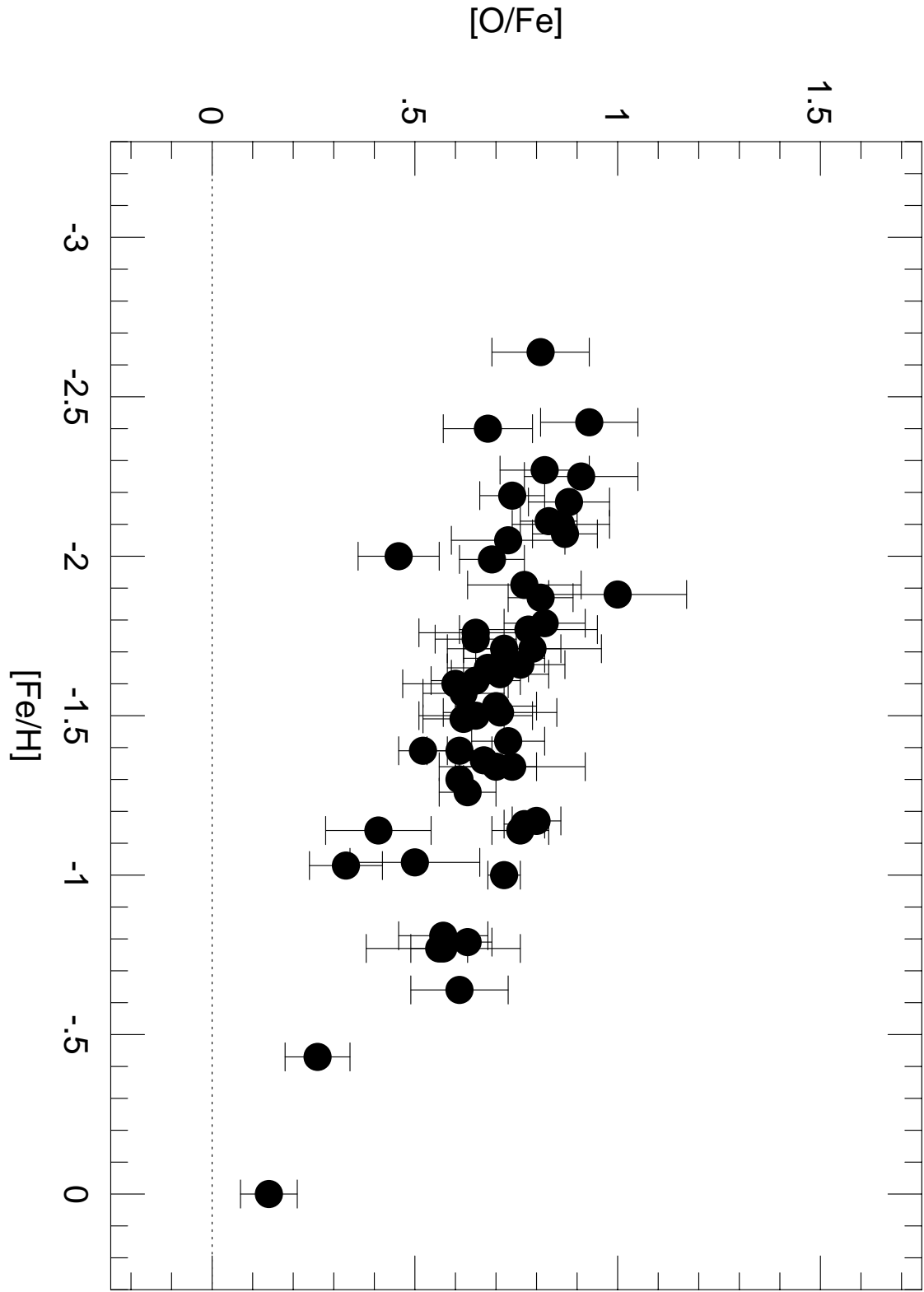}
\caption{The [O/Fe] vs. [Fe/H] diagram for the Ad Hoc scale.
Because the two oxygen indicators were forced to agree on this
scale, each star is indicated by a single point.  The error bars
are those derived for the forbidden lines, but the values derived
for the permitted lines are similar.}
\end{figure}

\clearpage
\begin{figure}[bht]
\plotone{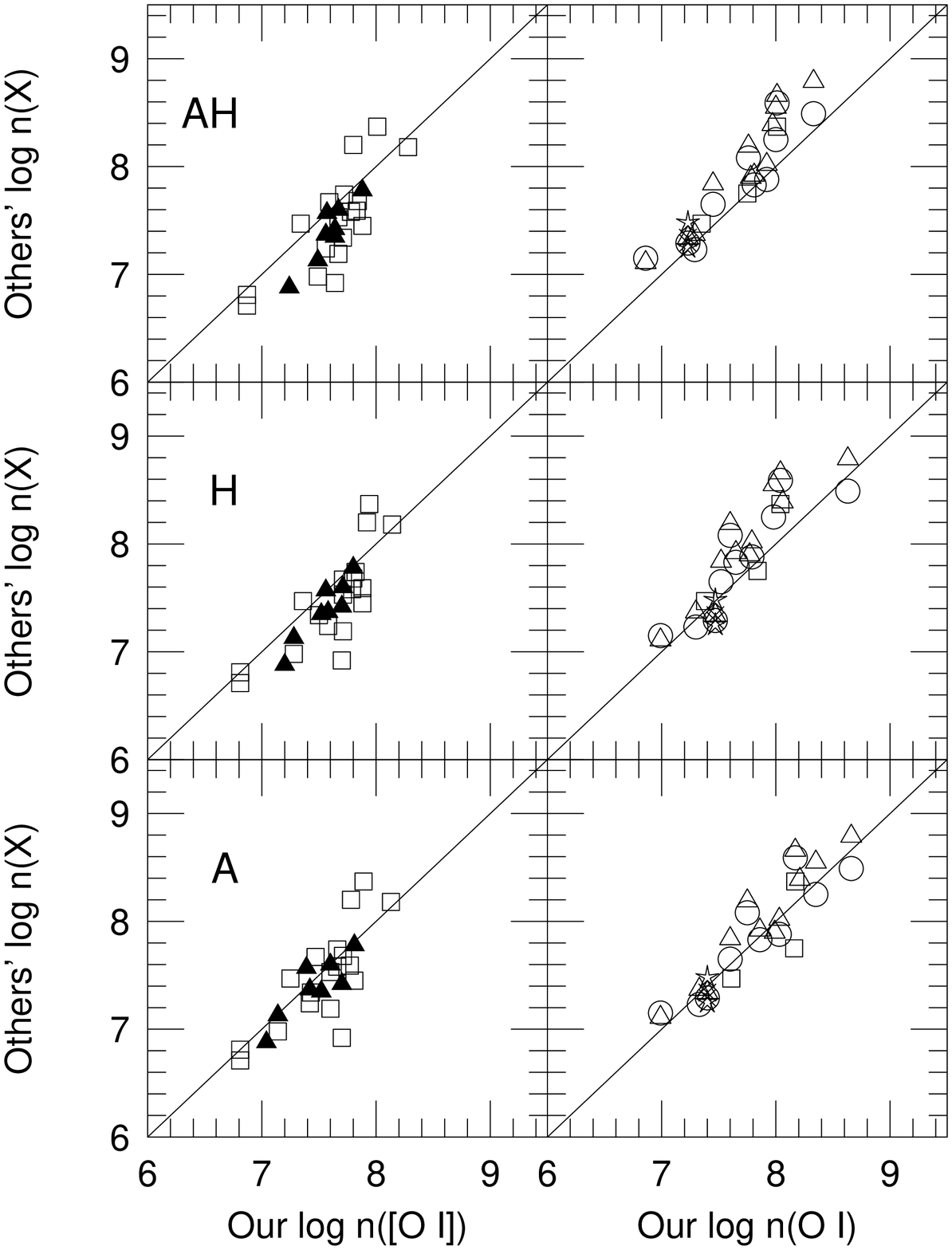}
\caption{Comparisons between the oxygen abundances
derived here and in other works.  Each row represents either the
Alsono (A), Houdashelt (H), or Ad Hoc (AH) scales, while the left
and right columns are for the forbidden and permitted line abundances,
repsectively.  The data are from \citealt{cg00} (squares),
\citealt{s96} (solid triangles), \citealt{m00} (circles), and
\citealt{cv97} (open triangles).  The stars represent the 
\citealt{i98} OH and \citealt{i01} O I abundances
for BD~+23~3130. }
\end{figure}

\clearpage
\begin{figure}[bht]
\plotone{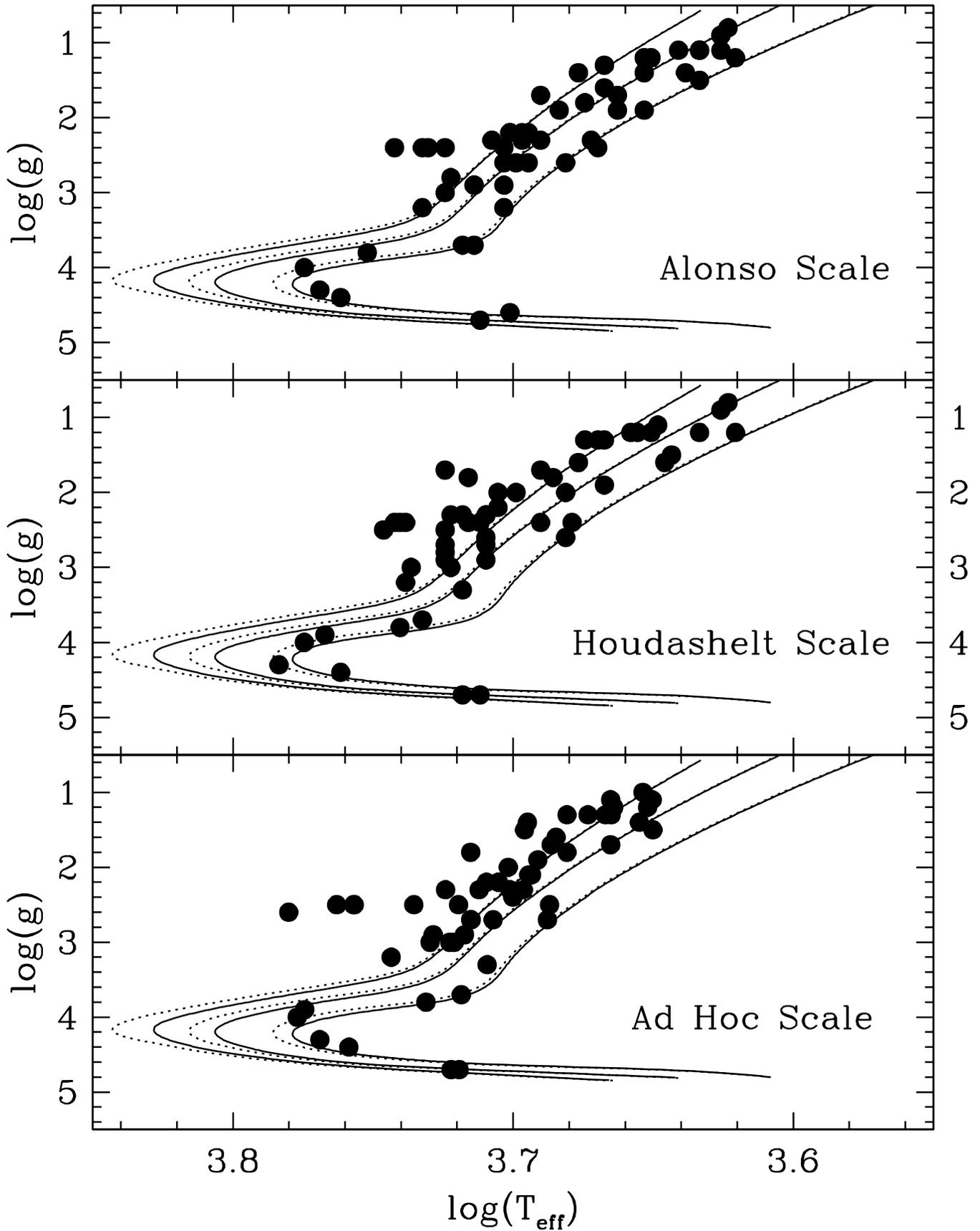}
\caption{The $\log{T_{eff}}$--$\log{g}$ plane for
the three parameter scales analyzed here. The lines are 10 (dotted) and
12 (solid) Gyr Vandenberg et al. (2000) isochrones of [Fe/H] = $-2.31$, $-1.54$, and $-0.84$ }
\end{figure}


\clearpage
\begin{figure}[bht]
\plotone{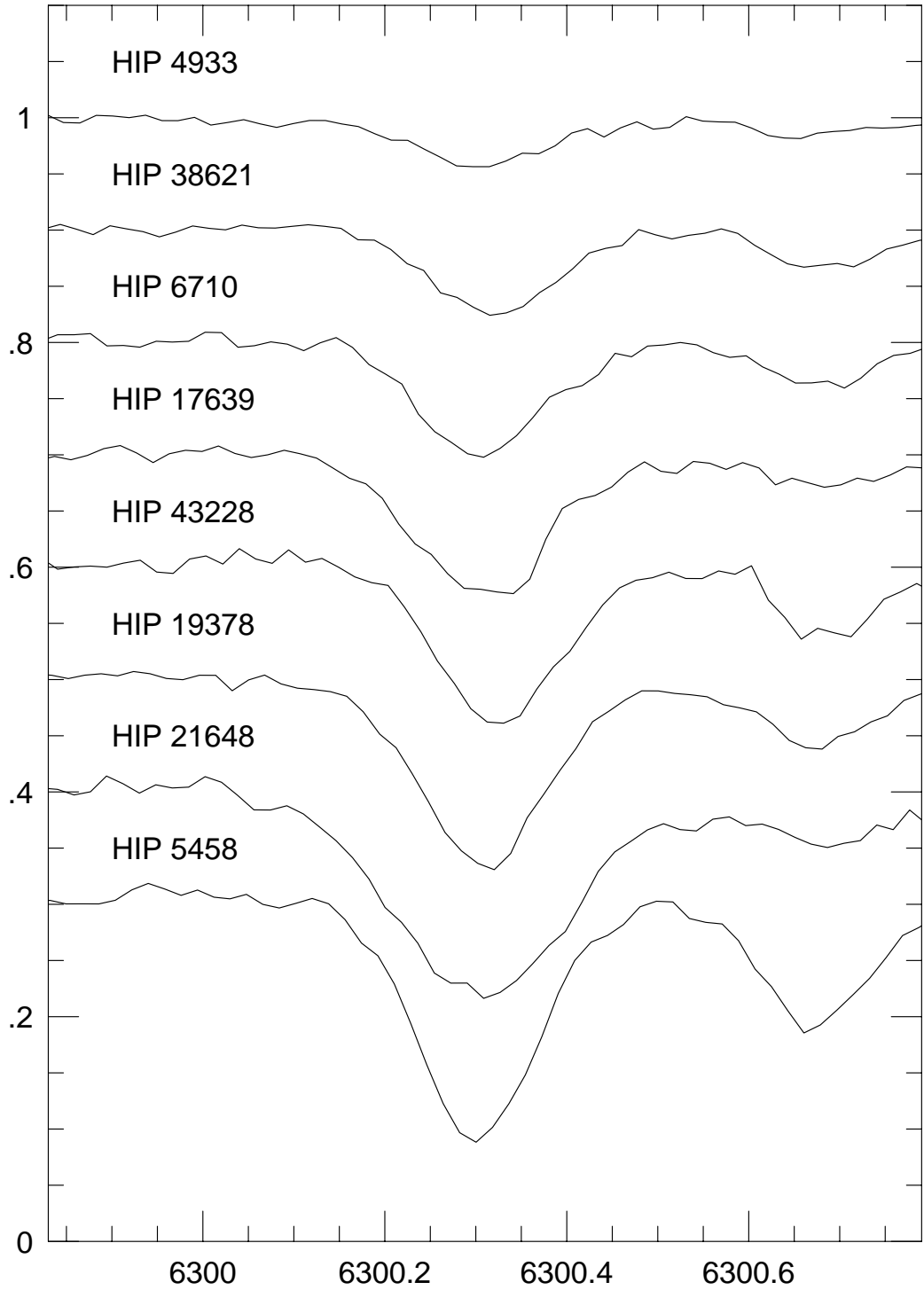}
\caption{The 6300.31 \AA{} region for eight giants. All
are on the same scale, but shifted vertically for clarity. None of the
lines here show any sign of emission features.}
\end{figure}
 
\clearpage
\begin{figure}[bht]
\plotone{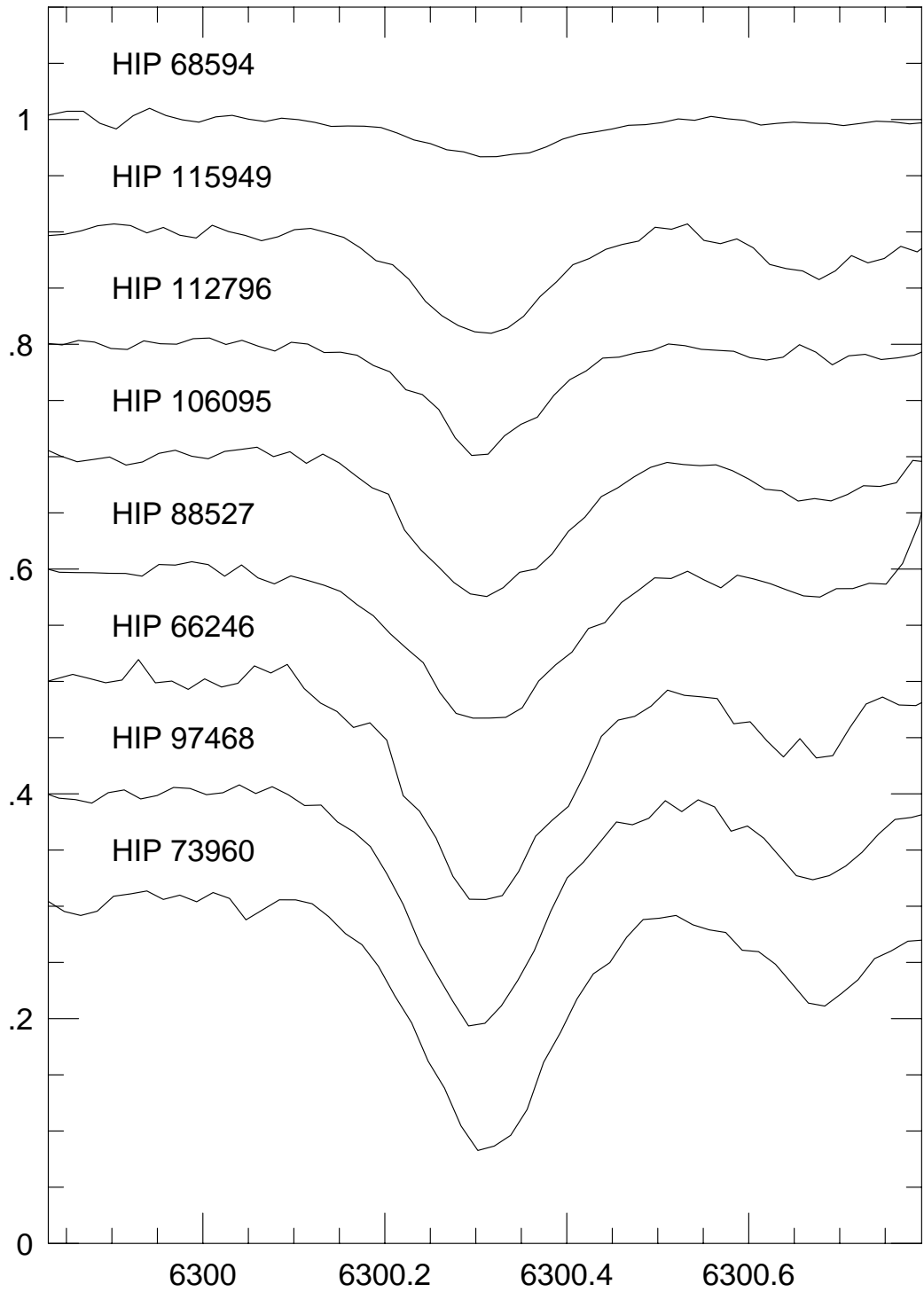}
\caption{Same as Figure 18 for eight more giants.}
\end{figure}

\end{document}